\documentclass[twocolumn]{revtex4}
\usepackage[utf8]{inputenc}
\usepackage{graphicx}

\begin{document}

\title{Electric polarization evolution equation for antiferromagnetic multiferroics
with the polarization proportional to the scalar product of the spins}

\author{Pavel A. Andreev}
\email{andreevpa@physics.msu.ru}
\affiliation{Department of General Physics, Faculty of physics, Lomonosov Moscow State University, Moscow, Russian Federation, 119991.}
\author{Mariya Iv. Trukhanova}
\email{trukhanova@physics.msu.ru}
\affiliation{Faculty of physics, Lomonosov Moscow State University, Moscow, Russian Federation, 119991.}
\affiliation{Russian Academy of Sciences, Nuclear Safety Institute (IBRAE), B. Tulskaya 52, Moscow, Russian Federation, 115191.}

\date{\today}

\begin{abstract}
The spin current model of electric polarization in multiferroics is justified
via the quantum hydrodynamic method and the mean-field part of the spin-orbit interaction.
The spin current model is applied to derive the electric polarization proportional to the scalar product of the spins of the nearby ions,
which appears to be caused by the Dzylaoshinskii-Moriya interaction.
The symmetric tensor spin structure of the polarization is discussed as well.
We start our derivations for the ferromagnetic multiferroic materials
and present further generalizations for the antiferromagnetic multiferroic materials.
We rederive the operator of the electric dipole moment,
which provides the macroscopic polarization obtained via the spin current model.
Finally, we use the quantum average of the found electric dipole moment operator to derive
the polarization evolution equation for the antiferromagnetic multiferroic materials.
The possibility of spiral spin structures is analyzed.
\end{abstract}

\maketitle

\section{Introduction}

The Landau--Lifshitz--Gilbert equation happens to be a highly effective method
for the theoretical study of the macroscopic processes of the magnetization evolution
in the magnetically ordered materials.
However, if we dial with the multiferroic materials,
it requires an equation for the polarization evolution.
Therefore, the problem of the derivation of the polarization evolution equation is formulated
\cite{Andreev 23 11}, \cite{Andreev 23 12}.
There are three mechanisms of the polarization formation in II-type multiferroics \cite{Tokura RPP 14},
where the magnetic properties related to the dielectric properties,
while the I-type multiferroics are the materials,
where the magnetic properties and the ferroelectric properties coexist without strong interference.
These mechanisms are the exchange-striction mechanism (symmetric-parallel components of spins gives the polarization),
the spin-current/inverse Dzylaoshinskii-Moriya model (antisymmetric-perpendicular components of spins gives the polarization),
the spin dependent p–d hybridization related to the spin of single magnetic ion.
Described mechanisms of ferroelectricity of spin origin are summarized in Fig. 2 of Ref. \cite{Tokura RPP 14}.

Particularly, there is the inverse Dzylaoshinskii-Moriya model associated with the spin current,
which is also called the spin current model.
In this paper, we are focused on the analytical justification of the spin current model,
which basically states that the polarization is proportional to the spin-current
$P^{\alpha}\sim\varepsilon^{\alpha\beta\gamma}J^{\beta\gamma}$,
where
$P^{\alpha}$ is the polarization,
$\varepsilon^{\alpha\beta\gamma}$ is the three-dimensional Levi-Civita symbol,
$J^{\beta\gamma}$ is the spin current tensor.
We show that it can be derived with no relation to the particular mechanism of the polarization formation.
Further application of the spin current caused by different mechanisms leads to
the electric dipole moment being proportional to the scalar product of the spins,
or
the electric dipole moment being proportional to the vector product of the spins.
The analytical derivation of polarization
(its macroscopic form and corresponding microscopic electric dipole moment)
allows to specify the interaction leading to the physical mechanisms of the polarization formation in two mentioned regimes.

The polarization evolution equation for multiferroics with ferromagnetic order of spins
for two of described mechanisms (the symmetric and antisymmetric regimes)
are considered in Refs. \cite{Andreev 23 11}, \cite{Andreev 23 12}.
Here, we consider the polarization evolution equation for the antiferromagnetic multiferroics,
where the electric polarization is proportional to the scalar product of the spins of the neighboring ions.

The Landau--Lifshitz--Gilbert
equation contains a number of terms representing major phenomena existing in the magnetically ordered materials.
One of the major interactions in the ferromagnetic materials is the exchange interaction,
which is modeled by the Heisenberg Hamiltonian.

Multiferroic materials are the magnetically ordered materials,
so similar mechanisms can affect the dynamics of polarization in the multiferroic materials.
It is especially related to the II-type multiferroics,
where the magnetic and the dielectric phenomena are deeply related,
in contrast to the I-type multiferroics,
where the magnetic and the dielectric phenomena coexist.
Therefore, we consider the role of exchange interaction in the evolution of polarization
in the regime,
where the electric dipole moment is proportional to the scalar product of the spins.

A classification of ferroelectrics is also presented in Ref. \cite{Cheong NM 07}.
It shows
the relation between materials and mechanism of the polarization formation
(mechanism of inversion symmetry breaking) in this material (see Table 1).
It shows five mechanisms with relation to particular materials,
but no analytical formalization is given in contrast with Ref. \cite{Tokura RPP 14}.
Nevertheless, Ref. \cite{Cheong NM 07} also includes the spin-atomic structure for a–b plane of perovskite YNiO$_{3}$ (see Fig. 1d),
where we see antiparallel spins for the net of Ni$^{3+\delta}$ (smaller value of magnetic moment),
we also see antiparallel spins for the net of Ni$^{3-\delta}$ (larger value of magnetic moment),
while pairs of nearest Ni$^{3+\delta}$ and Ni$^{3-\delta}$ have parallel spins.
This antiferromagnetic system of parallel/antiparallel spins can be associated with the results obtained in this paper.
The fourth-order term in energy $\sim - P^{2}M^{2}$ is associated with YMnO$_{3}$ and BiMnO$_{3}$ \cite{Cheong NM 07}
(see the text before eq. 1 on page 16).
It corresponds to the structural transition "Geometric ferroelectrics."
It can be associated with the homogeneous magnetic ordering \cite{Cheong NM 07}.

In contrast, the polarization can be described by the following equation for the inhomogeneous magnetic ordering
\cite{Baryachtar JETP Lett 83}, \cite{Stefanovskii Sov. J. Low Temp. Phys 86}, \cite{Mostovoy PRL 06}
\begin{equation}\label{MFMafmUEI P by Mostovoy}
\textbf{P}\sim (\textbf{S}\cdot\nabla)\textbf{S}-\textbf{S}(\nabla\cdot\textbf{S})
, \end{equation}
where the inhomogeneous magnetization can be associated with the rotation of the magnetic moments
(see Figs. 4 and 5 in Ref. \cite{Cheong NM 07}).
The antiferromagnetic analog of eq. (\ref{MFMafmUEI P by Mostovoy}) can be found in Refs.
\cite{Zvezdin JETP Lett 09}, \cite{Sosnowska JMMM 95}.

A model of the microscopic origin of electric polarization is given in Ref. \cite{Solovyev PRL 21}.
The focus is made on the noncollinear magnetic order and formation of the electric polarization in Mott insulators.
However, it is also shown that the form of the magnetoelectric coupling allows additional constructions in comparison with known configurations \cite{Tokura RPP 14}.

A slow change/rotation of the spin direction in a sample
(see Fig. 4 in Ref. \cite{Cheong NM 07})
creates conditions for the noncollinear mechanisms of the polarization formation
(see Fig. 2 (d)–(f) of Ref. \cite{Tokura RPP 14}).
However, the collinear part of the relative spin orientation is nonzero,
so the mechanism of the polarization related to the collinear spins can contribute in these systems.
Coefficients giving the polarization can differs for different mechanisms,
so one of mechanisms can be suppressed.

The area of the spin rotation leading to the noncollinear and the collinear spin formations appears in the domain walls.
Magnon-induced domain wall motion in ordinary ferromagnets is considered in Ref. \cite{Mikhailov JETP Lett 84}.
The Dzyaloshinskii-Moriya interaction as a mechanism of the magnon-driven domain-wall motion is considered in Ref. \cite{Wang PRL 15}.
Antiferromagnetic domain wall motion induced by spin waves is considered in Ref. \cite{Tveten PRL 14}.
Magnon induced magnetization dynamics in multiferroics is considered in Ref. \cite{Risinggard SR 16}.
It includes the magnon-induced domain wall motion.
Particularly, the Landau–-Lifshitz–-Gilbert equation is applied for the analysis,
where an additional term is included to describe the coupling between the electric field and the magnetization
in a manner associated with eq. (\ref{MFMafmUEI P by Mostovoy}) (see equations 3 and 4).
Magnonic spin-transfer torque can
efficiently drive a domain wall to propagate in the opposite direction to that of the spin wave,
as demonstrated in Ref. \cite{Yan PRL 11}.
An analytical derivation of the magnon-driven Dzyaloshinskii-Moriya torque is developed in Ref. \cite{Manchon PRB 14}.
It is also shown that the Dzyaloshinskii-Moriya interaction
mediated by spin waves can generate a torque on a homogeneous magnetization that resembles the Rashba torque.
The magnetization dynamics in a thin-film ferromagnet deposited on a topological insulator is studied \cite{Linder PRB 14}.
It is focused on domain-wall motion via current and the possibility of a spin-wave torque acting on the magnetization.
The coupling between the magnetic
domain wall and the topological insulator removes the degeneracy of the wall profile
with respect to its chirality and topological
charge, as shown analytically \cite{Linder PRB 14}.
The dynamics of a multiferroic domain wall in which an electric field can couple to the magnetization
via the inhomogeneous magnetoelectric interaction is studied in Ref. \cite{Chen JAP 14}.
It also demonstrates that in the stationary regime, the chirality of the domain wall can be efficiently
reversed when the electric field is applied along the direction of the magnetic field.
Some discussion of the structure of the Dzyaloshinskii constant and the Dzyaloshinskii-Moriya interaction can be found in
Ref. \cite{Chen PRL 13} in context of the experimental analysis of magnetized Fe/Ni bilayers,
where a new type of domain wall
structure is reported.

Another object with the noncollinear spin structure is Skyrmions.
Let us mention that the skyrmion-like spin structures
formed by the thermal fluctuations in the ferromagnetic materials are studied in \cite{Huang PRL 12}.
In Ref. \cite{Huang PRL 12}
it is also specified that
the uniaxial anisotropy $K$
($K>0$, favoring perpendicular magnetic anisotropy with an energy of $-KM_z^2$)
and "effective stiffness" $K_0$
of the conical phase can be calculated from the in-plane
and out-of-plane saturation fields.

In this paper we mainly consider the single-phase AFM multiferroics.
However, important novel phenomena are discovered in multilayered,
where the film is grown on the substrate.
Application of our model depends on the properties of the materials.
The current form of our model includes the connection between magnetic and dielectric properties of the materials,
while multiferroics are well-known due to relation of the magnetic and dielectric properties to the elastic properties of the medium.
However, the elasticity has not been considered in terms of our model.
As an illustration we mention AFM/ferroelectric multiferroic heterostructures,
such as experimentally demonstrated Mn-Pt/PMN-PT heterostructures \cite{Liu NE 18}.
Application of the electric field to BaTiO$_3$ leads to the strain appearance.
As the consequence the strain appears in the intermetallic Mn$_3$Pt film.
It causes changes in the magnetic properties of Mn$_3$Pt film.
Particularly it changes properties of the topological anomalous Hall effect in a non-collinear phase of the antiferromagnet.
So, the elastic properties play essential role in this phenomenon.
Another example is the tunneling magnetoresistance effect \cite{Qin N 23}.
An effect similar to the ferromagnet-antiferromagnet exchange-bias system is discovered
for a room-temperature exchange-bias effect between a collinear antiferromagnet and a non-collinear antiferromagnet.
The N$\acute{e}$el pair anisotropy term between the nearest Mn-Pt pairs
(equation 1 in Sec. Methods of Supplementary materials of Ref. \cite{Qin N 23}) is applied at the numerical modeling.
It structure is similar to the polarization in the spin dependent p-d hybridization mechanism.

Describing the magnetization and its dynamics,
we usually focused on the spin density,
while the spin of atom/ion is formed by the spins and orbital motion of elements of the atomic structure.
The superposition of the spin angular momentum density
and the orbital angular momentum density is obtained in Ref. \cite{Yan PRB 13} (see Eq. (14)).
The conservation of the $z$ component of the total angular momentum is demonstrated
for a Heisenberg ferromagnet with isotropic exchange interaction \cite{Yan PRB 13}.

We consider the Dzyaloshinskii-Moriya interaction to complete the spin current model for collinear spins
(it is assumed that the collinear spin structure is formed by the Heisenberg exchange interaction,
while further formation of spin related electric polarization is formed by the Dzyaloshinskii-Moriya interaction).
So, we want to mention some continuous approaches based on the Dzyaloshinskii-Moriya Hamiltonian.
The model used in Ref. \cite{Wang PRL 15}
is based on the total free energy with a nonexplicit account of the Dzyaloshinskii-Moriya interaction.
Its explicit contribution is given in the Landau-Lifshitz-Gilbert equation in the form of the torque
$\textbf{T}=\gamma \tilde{D}\textbf{M}\times(\nabla\times \textbf{M})$,
with $\gamma$ is the gyromagnetic ratio.
It corresponds to the Dzyaloshinskii-Moriya interaction with
a chiral energy of $\tilde{D}\textbf{M}\cdot(\nabla\times \textbf{M})$.
In our work, we address the Dzyaloshinskii-Moriya interaction given by Hamiltonian
$\hat{H}=(-1/2)\textbf{D}_{ij}\cdot
[\hat{\textbf{s}}_{i}\times\hat{\textbf{s}}_{j}]$
and derive a different form of the torque in the Landau-Lifshitz equation
for the ferromagnetic and antiferromagnetic materials.
In our derivation, we include the vector nature of the Dzyaloshinskii constant and its analytical structure.

The Dzyaloshinskii-Moriya interaction generates the torque on the magnetization.
The Dzyaloshinskii-Moriya interaction mediated by spin waves is considered in Ref. \cite{Manchon PRB 14}
for systems displaying the interfacial Dzyaloshinskii-Moriya interaction.
A thin magnetic film with the magnetization $\textbf{m}$ aligned along the in-plane easy axis,
the magnetic energy associated with Dzyaloshinskii-Moriya interaction reads \cite{Manchon PRB 14}
$W=-D \textbf{m} \times[(\textbf{z}\times\nabla)\times \textbf{m}]$,
where
system with an interfacial inversion asymmetry along the normal $z$
\cite{Bogdanov PRL 01}
(films and multilayers with in-plane and out-of-plane magnetization are considered with the prediction of the
two-dimensional localized patterns),
\cite{Thiaville EPL 12}
(domain walls in ultrathin magnetic films are considered there).
It leads to an effective fieldlike torque of the form $T\sim (\textbf{m} \times \textbf{z})\times \textbf{j}_{m}$,
with $\textbf{j}_{m}$ is the spin-wave current.

The many-particle quantum hydrodynamic method has been developed for the structureless mediums
such as the quantum ultracold gases \cite{Andreev LP 21 fermions}, \cite{Andreev PoF 21}
and plasma-like mediums 
However, it has been shown that it is possible to capture some features of solid state.
Particularly, the material field form of the Landau––Lifshitz equation is derived
from the many-particle Pauli equation in the coordinate representation \cite{Andreev Trukhanova Vestn 23}.
It opened a possibility for the anlysis of the multiferroic materials in terms of the quantum hydrodynamic method.
Therefore, in order to study the antiferromagnets,
we present the derivation of the antiferromagnet analog of the Landau––Lifshitz equation.
Next, in this paper, we develop a microscopically justified macroscopic spin current model.
We also derive and apply the spin current caused by the Dzylaoshinskii-Moriya interaction to complete the derivation of the polarization.
Obtained macroscopic polarization is compared with earlier suggested electric dipole moments of the pair of magnetic ions \cite{Tokura RPP 14}.
The polarization evolution equation is derived for the found form of polarization caused by the Dzylaoshinskii-Moriya interaction.

This paper is organized as follows.
In Sec. II the microscopic derivation of the spin evolution equation for the antiferromagnetic materials is developed
within the many-particle quantum hydrodynamic method.
In Sec. III the approximate form of the polarization is considered for ferromagnetic and antiferromagnetic multiferroics,
if the electric dipole moment is proportional to the scalar product of the spins.
In Sec. IV the spin current model is derived from the momentum balance equation with the spin-orbit interaction.
Furthermore, in Sec. IV, the spin-current caused by the Dzylaoshinskii-Moriya interaction is presented in order to obtain corresponding polarization.
In Sec. V the polarization evolution equation is derived as the evolution of the quantum average of the electric dipole moment operator
under the influence of the Zeeman energy and the Coulomb exchange interaction,
both for ferromagnetic and antiferromagnetic multiferroics.
In Sec. VI the equilibrium solutions for the obtained model of multiferroics are discussed, including the spiral structures.
In Sec. VII a brief summary of obtained results is presented.

\section{The spin evolution equation for the antiferromagnetic material:
the microscopic--based derivation with the exchange interaction}

If we consider the spin evolution equation in the ferromagnetic materials with primarily exchange interaction,
we find $\partial_{t}\textbf{S}=(1/6)g_{u}[\textbf{S},\triangle\textbf{S}]$,
where $g_{u}=\int \xi^{2}U(\xi)d^{3}\xi$.
Firstly, it is based on the Heisenberg Hamiltonian
$\hat{H}= -\frac{1}{2}\sum_{i=1}^{N}\sum_{j=1, j\neq i}^{N} U_{1}(r_{ij})
(\hat{\textbf{s}}_{i}\cdot\hat{\textbf{s}}_{j})$.
Function $U_{1}(r_{ij})$ is the exchange integral.
It depends on the distance between interaction particles.
In combination with the spin operators,
it gives an effective potential energy.
Function $U_{1}(r_{ij})$ drops at the distance larger average interparticle distance.
So, it includes the interaction of neighboring atoms or ions.
It can include the interaction of the atoms separated by an atom,
but it does not include the influence of atoms located at the further distances.

In order to get a systematic derivation of the spin evolution equation,
we need to define the spin density of the system.
For the quantum systems,
it is defined as the quantum average of the spin operator $\hat{\textbf{s}}_{i}$
\begin{equation}\label{MFMafmUEI S def} \textbf{S}(\textbf{r},t)=
\int \Psi^{\dagger}(R,t)\sum_{i}\delta(\textbf{r}-\textbf{r}_{i})
(\hat{\textbf{s}}_{i}\Psi(R,t))dR, \end{equation}
where $i$ is the number of atoms.
The spin operators obey the commutation relation
\begin{equation}\label{MFMafmUEI commutator of spins}
[\hat{s}_{i}^{\alpha},\hat{s}_{j}^{\beta}]
=\imath\hbar\delta_{ij}\varepsilon^{\alpha\beta\gamma} \hat{s}_{i}^{\gamma}, \end{equation}
where
$\alpha$, $\beta$, $\gamma$ are the tensor indexes,
so each of them is equal to $x$, $y$, $z$,
summation on the repeating Greek symbol is assumed,
$\imath$ is the imaginary unit,
$\delta_{ij}$ is the three-dimensional Kronecker symbol,
$\varepsilon^{\alpha\beta\gamma}$ is the three-dimensional Levi-Civita symbol.

We consider systems of atoms or ions.
Hence, we deal with structured objects or particles.
There is the exchange interaction of valence electrons in each ion.
It makes a contribution to the properties at the ion as the particle under consideration.
However, there is a short-range correlation between neighboring ions due to the exchange interaction of valence electrons belonging to different ions.
This interaction is included in the model presented below via the Heisenberg Hamiltonian.

Evolution of the wave function of the system of ions
is described by the Pauli equation
\begin{equation}\label{MFMafmUEI Sch eq}
\imath\hbar\partial_{t}\Psi(R,t)
=\hat{H}\Psi(R,t).\end{equation}
In this paper, we mostly focused on the evolution of spin density and polarization
under influence of the Zeeman energy and the Coulomb exchange interaction
\begin{equation}\label{MFMafmUEI H Ham short HH and mB}
\hat{H}= -\sum_{i=1}^{N}\hat{\mbox{\boldmath $\mu$}}_{i}\cdot\textbf{B}_{i}
-\frac{1}{2}\sum_{i=1}^{N}
\sum_{j=1, j\neq i}^{N} U(r_{ij})(\hat{\textbf{s}}_{i}\cdot\hat{\textbf{s}}_{j}), \end{equation}
$\hbar$ is the Planck constant,
$N$ number of atoms/ions in the system,
$\Psi(R,t)$ many-particle wave function of the system,
$R=\{\textbf{r}_{1}, ..., \textbf{r}_{N}\}$,
$\textbf{B}_{i}$ is the magnetic field,
acting on $i$-th atom,
$\hat{\mbox{\boldmath $\mu$}}_{i}$ magnetic moment operator,
which is proportional to the spin operator
$\hat{\mbox{\boldmath $\mu$}}_{i}=\gamma_{i}\hat{\textbf{s}}_{i}$
with the gyromagnetic ratio $\gamma_{i}$,
$U_{ij}=U(\textbf{r}_{i}-\textbf{r}_{j})$
is the exchange integral of the Heisenberg Hamiltonian as the function of the interparticle distance,
(the exchange integral).
Let us repeat that $N$ is the full number of atoms under consideration.
It can be considered as the sum of numbers of particles in each of two species of the magnetic atoms $N=N_{A}+N_{B}$.
Formally, we have interaction between all pairs of atoms in the system in the second term of Hamiltonian (\ref{MFMafmUEI H Ham short HH and mB}),
but the short range of function $U(r_{ij})$ leads to the contribution of the neighboring atoms only.

If we consider two subsystems for the antiferromagnetic material,
we need to define the spin density for each subsystem
\begin{equation}\label{MFMafmUEI S def} \textbf{S}_{s}(\textbf{r},t)=
\int \Psi^{\dagger}(R,t)\sum_{i\in s}\delta(\textbf{r}-\textbf{r}_{i})
(\hat{\textbf{s}}_{i}\Psi(R,t))dR, \end{equation}
where index $s$ refers to the number of the species of atoms,
or, in the simplest case,
it can be atoms of the same isotope with opposite spin projection.
We focus on the system of two subspecies $s=A$ and $s=B$.

Next, we consider the spin evolution equation.
The time derivative acts on the wave function,
while the time derivative of the wave function is replaced with a Hamiltonian in accordance with the Pauli equation
\begin{equation}\label{MFMafmUEI S time der via H} \partial_{t}\textbf{S}_{s}(\textbf{r},t)=
\frac{\imath}{\hbar}\int \sum_{i\in s}\delta(\textbf{r}-\textbf{r}_{i})\Psi^{\dagger}(R,t)
[\hat{H},\hat{\textbf{s}}_{i}]\Psi(R,t)dR. \end{equation}

The partial contribution in the spin evolution equation from the Zeeman energy
$-\sum_{i=1}^{N}\hat{\mbox{\boldmath $\mu$}}_{i}\cdot\textbf{B}_{i}$
in Hamiltonian (\ref{MFMafmUEI H Ham short HH and mB}) leads to
$\partial_{t}\textbf{S}_{s}=\frac{2\mu}{\hbar}[\textbf{S}_{s},\textbf{B}]$.
If we consider the interaction of the nearest neighbors,
which corresponds to the interaction of the different subspecies,
we find the following contribution of the second term of Hamiltonian (\ref{MFMafmUEI H Ham short HH and mB})
in the spin evolution equation
\begin{equation}\label{MFMafmUEI s evolution HH AB}
\partial_{t}\textbf{S}_{s}=
g_{0u,AB}[\textbf{S}_{s},\textbf{S}_{s'\neq s}]
+\frac{1}{6}g_{u,AB}[\textbf{S}_{s},\triangle\textbf{S}_{s'\neq s}]. \end{equation}

If we consider the interaction
second row neighbors,
which corresponds to the interaction of atoms of the same species
we find
\begin{equation}\label{MFMafmUEI s evolution HH AA}
\partial_{t}\textbf{S}_{s}=\frac{1}{6}g_{u}[\textbf{S}_{s},\triangle\textbf{S}_{s}]. \end{equation}

If we consider antiferromagnetic material composed of atoms of the same species,
we have same form of potential of interaction between atoms of the same subspecies and atoms of different subspecies.
However, the signs of the potentials are different
since the antiferromagnetic order corresponds to the negative exchange integral $g_{u,AB}<0$
while atoms with the same spin direction have a positive exchange integral $g_{u}>0$.
So we can assume the following relation between the interaction constants $g_{u,AB}=-g_{u}$.
Similarly, we have a relation for the zeroth order constant $g_{0u,AB}=-g_{0u}$.
It allows us to combine equations (\ref{MFMafmUEI s evolution HH AB}) and (\ref{MFMafmUEI s evolution HH AA})
in order to get the spin evolution equation under the exchange interaction of two subspecies
$$\partial_{t}\textbf{S}_{s}=
\frac{2\mu}{\hbar}[\textbf{S}_{s},\textbf{B}]
-g_{0u}[\textbf{S}_{s},\textbf{S}_{s'\neq s}]$$
\begin{equation}\label{MFMafmUEI s evolution HH AA and AB}
+\frac{1}{6}g_{u}[\textbf{S}_{s},\triangle(\textbf{S}_{s}-\textbf{S}_{s'\neq s})]. \end{equation}
It is well known that the modeling of the antiferromagnetic materials includes superpositions of the partial magnetizations \cite{Landau vol 9},
in our case spin densities.
Hence, we introduce $\mbox{\boldmath $\Sigma$}=\textbf{S}_{A}+\textbf{S}_{B}$
and $\textbf{L}=\textbf{S}_{A}-\textbf{S}_{B}$.
In literature, $\textbf{L}$ is used for the difference of the magnetizations \cite{Landau vol 9},
hopefully it would not confuse the readers.
It leads to the following equations
\begin{equation}\label{MFMafmUEI s evolution HH Sigma}
\partial_{t}\mbox{\boldmath $\Sigma$}=
\frac{2\mu}{\hbar}[\mbox{\boldmath $\Sigma$},\textbf{B}]
+\frac{1}{6}g_{u}[\textbf{L},\triangle\textbf{L}], \end{equation}
and
\begin{equation}\label{MFMafmUEI s evolution HH L}
\partial_{t}\textbf{L}=
\frac{2\mu}{\hbar}[\textbf{L},\textbf{B}]
+\frac{1}{6}g_{u}[\mbox{\boldmath $\Sigma$},\triangle\textbf{L}]
+g_{0u,AB}[\textbf{L},\mbox{\boldmath $\Sigma$}], \end{equation}
where
we used simple representation
$[\textbf{S}_{A},\textbf{S}_{B}]$
$=[\textbf{S}_{A}-\textbf{S}_{B},\textbf{S}_{B}]$
$=[\textbf{S}_{A}-\textbf{S}_{B},\textbf{S}_{A}+\textbf{S}_{B}-\textbf{S}_{A}]$
$=[\textbf{L},\mbox{\boldmath $\Sigma$}]$
$-[\textbf{S}_{A}-\textbf{S}_{B},\textbf{S}_{A}]$
$=[\textbf{L},\mbox{\boldmath $\Sigma$}]$
$-[\textbf{S}_{A},\textbf{S}_{B}]$.
So, we get
$2[\textbf{S}_{A},\textbf{S}_{B}]$$=[\textbf{L},\mbox{\boldmath $\Sigma$}]$.
Here we have $\mid\textbf{S}_{A}\mid\approx\mid\textbf{S}_{B}\mid$,
and $\textbf{S}_{A}\approx-\textbf{S}_{B}$.
So, the sum of partial spin densities
$\mbox{\boldmath $\Sigma$}=\textbf{S}_{A}+\textbf{S}_{B}$ is a small value in the antiferromagnetic material,
and $\mid\textbf{L}\mid\approx2\mid\textbf{S}_{A}\mid\approx2\mid\textbf{S}_{B}\mid$.

Equation (\ref{MFMafmUEI s evolution HH Sigma}) is obtained for a small vector $|\mbox{\boldmath $\Sigma$}|\ll |\textbf{L}|$.
The first (second) term on the right-hand side is proportional
to the small vector $\mbox{\boldmath $\Sigma$}$
(to the small combination of parameters $g_{u}\triangle\textbf{L}$).
In equation (\ref{MFMafmUEI s evolution HH L})
we see
the first term with no small parameters,
the third term containing the small vector $\mbox{\boldmath $\Sigma$}$,
and the second term containing the product of the small parameters $\mbox{\boldmath $\Sigma$}$ and $g_{u}\triangle\textbf{L}$.
So, the second term can be dropped in further applications.

Equations (\ref{MFMafmUEI s evolution HH Sigma}) and (\ref{MFMafmUEI s evolution HH L})
are well-known for the antiferromagnetic materials \cite{Landau vol 9}.
However, our derivation allows us to establish the explicit form of coefficients in this equation
in relation to the microscopic nature of the interaction entering the Pauli equation (\ref{MFMafmUEI H Ham short HH and mB}).
Our major goal in this paper is the derivation of the polarization evolution equation for the multiferroic materials.
However, the derivation of equations
(\ref{MFMafmUEI s evolution HH Sigma}) and (\ref{MFMafmUEI s evolution HH L})
shows the usefulness of our method of derivation of the macroscopic equations from the microscopic theory.
Details of the derivation are not considered here,
but the method of derivation can be found in Ref. \cite{Andreev 23 12} and \cite{Andreev LP 21 fermions}.
This work on the microscopic justification of the macroscopic equations for the magnetization and the polarization
is a part of an interdisciplinary field,
where similar justifications are made for the classical and quantum systems
\cite{Andreev LP 21 fermions}, \cite{Andreev PoF 21}, \cite{Andreev Trukhanova Vestn 23},
\cite{Andreev JPP 24}, \cite{Andreev 22 12}.

In this paper we focused on a partial microscopic derivation of the Landau--Lifshitz--Gilbert equation.
Here we presented the contribution of the Heisenberg Hamiltonian.
Below we consider the Zeeman energy and the Dzylaoshinskii-Moriya interaction.
We also consider the spin-orbit interaction as the key element for the spin current model.
However, the anisotropy energy and the Gilbert damping are consciously ignored,
while these terms are crucial for the complete model of the magnetically ordered materials.

\section{Microscopic definition of the macroscopic polarization in the exchange striction model}

The electric dipole moment related to a pair of neighboring magnetic ions
in the exchange striction regime is proportional to the scalar product of spins of these ions
\cite{Tokura RPP 14},
\cite{Aguilar PRL 09},
\cite{Miyahara arxiv 08}
\begin{equation}\label{MFMafmUEI edm def i i+1}
\textbf{d}_{i}\sim \mbox{\boldmath $\Pi$}(\textbf{s}_{i}\cdot\textbf{s}_{i+1}).\end{equation}
The nonmagnetic ions contributing to the dipole moment are not considered explicitly in this equation.
Equation (\ref{MFMafmUEI edm def i i+1}) is useful to analyze a linear structure.
Hence, we give the representation to the electric dipole moment of two ions
\begin{equation}\label{MFMafmUEI edm def simm}
d_{ij}^{\alpha}= \Pi_{ij}^{\alpha}(r_{ij})(\textbf{s}_{i}\cdot\textbf{s}_{j}), \end{equation}
where
$r_{ij}=\mid \textbf{r}_{i}-\textbf{r}_{j}\mid$ is the interparticle distance,
and dependence of vector function $\Pi_{ij}^{\alpha}(r_{ij})$ ensures
that the nearest neighbors give contribution in dipole moment $d_{ij}^{\alpha}$.

We develop the quantum theory of multiferroics,
so we need to make a transition to the operator form of the electric dipole moment
via the consideration of the spin operators
$\hat{d}_{ij}^{\alpha}= \Pi_{ij}^{\alpha}(\hat{\textbf{s}}_{i}\cdot\hat{\textbf{s}}_{j})$.
Considering all nearest neighbors of the $i$-th atom/ion,
we get the full electric dipole moment related to this atom
\begin{equation}\label{MFMafmUEI edm operator Mod}
\hat{d}_{i}^{\alpha}=\sum_{j\neq i}
\Pi_{ij}^{\alpha}(r_{ij})(\hat{\textbf{s}}_{i}\cdot\hat{\textbf{s}}_{j}). \end{equation}

We can use the operator of the electric dipole moment (\ref{MFMafmUEI edm operator Mod})
in order to find an approximate representation
via the spin density,
which can be an analog of the M. Mostovoy \cite{Mostovoy PRL 06},
but for different physical regime.
Therefore, we present the quantum average of operator $\hat{\textbf{d}}_{i}$,
which gives the macroscopic polarization
\begin{equation}\label{MFMafmUEI P def} \textbf{P}(\textbf{r},t)=
\int \Psi^{\dagger}(R,t)\sum_{i}\delta(\textbf{r}-\textbf{r}_{i})
(\hat{\textbf{d}}_{i}\Psi(R,t))dR. \end{equation}

Substituting operator (\ref{MFMafmUEI edm operator Mod})
in the polarization definition (\ref{MFMafmUEI P def})
and account of the formation of the electric dipole moment by the nearest neighbors allows to get the required approximate form of the polarization.
We need to explicitly introduce the interparticle distance in all functions in definition (\ref{MFMafmUEI P def})
as follows:
$\textbf{r}_{i}=\textbf{R}_{ij}+(1/2)\textbf{r}_{ij}$,
and
$\textbf{r}_{j}=\textbf{R}_{ij}-(1/2)\textbf{r}_{ij}$,
where
$\textbf{R}_{ij}=(\textbf{r}_{i}+\textbf{r}_{j})/2$,
and
$\textbf{r}_{ij}=\textbf{r}_{i}-\textbf{r}_{j}$.
This substitution includes the arguments of the wave function
$\Psi(R,t)=\Psi(...,\textbf{r}_{i}, ..., \textbf{r}_{j}, ...,t)$.

\subsection{The polarization of the ferromagnetic multiferroics}

In this paper,
we mostly focused on the antiferromagnetic materials.
However, for comparison,
it can be useful to obtain the polarization of the ferromagnetic multiferroics
in the "exchange striction" regime \cite{Tokura RPP 14}.
We use equation (\ref{MFMafmUEI P def}) with the operator of the electric dipole moment (\ref{MFMafmUEI edm operator Mod})
and make the expansion on the interparticle distance
(some details of the method are described in Ref. \cite{Andreev 23 12}).
It allows us to get an approximate expression of the polarization (\ref{MFMafmUEI P def}) in terms of the spin density:
\begin{equation}\label{MFMafmUEI P def expanded FR}
P^{\alpha}=
g_{0\Pi}^{\alpha} \textbf{S}^{2}
+\frac{1}{6}g_{\Pi}^{\alpha}(\textbf{S}\cdot\triangle \textbf{S}), \end{equation}
where
$g_{0\Pi}^{\alpha}=\int\Pi^{\alpha}(r)d\textbf{r}$,
and
$g_{(\Pi)}^{\alpha}=\int \xi^{2}\Pi^{\alpha}(\xi) d\mbox{\boldmath $\xi$}$.
Equation (\ref{MFMafmUEI P def expanded FR}) can be also represented in the following form
$P^{\alpha}=g_{0\Pi}^{\alpha} \textbf{S}^{2}
+\frac{1}{3}\frac{1}{2^{2}}g_{\Pi}^{\alpha}[\triangle \textbf{S}^{2} -2 (\partial_{\mu} S^{\nu})(\partial_{\mu} S^{\nu})]$.

Equation (\ref{MFMafmUEI P def expanded FR}) is the result of the expansion on the relative distance.
We included three terms of expansion.
The first and major term is proportional to $g_{0\Pi}^{\alpha}$.
The second term is equal to zero.
The third term is proportional to $g_{(\Pi)}^{\alpha}$ appears as a correction.

\subsection{The polarization of the antiferromagnetic multiferroics}

Formation of dipole (\ref{MFMafmUEI edm def i i+1}) happens due to the presence of the nonmagnetic ion
with the opposite charge between ions $i$ and $i+1$.
If we consider the antiferromagnetic material,
we find the alternation of the "spin-up" and "spin-down" ions.
So, if ion $i$ is "spin-up" then ion $i+1$ is "spin-down".
It shows that
if we consider a line of magnetic ions,
we have the nonmagnetic ion after each magnetic ion.
It has no relation to the spin direction of the magnetic ion.
Hence, the effective dipole moment (\ref{MFMafmUEI edm operator Mod})
can be associated with each magnetic ion ("spin-up" or "spin-down")
\begin{equation}\label{MFMafmUEI edm operator Mod s}
\hat{d}_{i\in s}^{\alpha}=\sum_{j\in s'}
\Pi_{ij}^{\alpha}(r_{ij})(\hat{\textbf{s}}_{i}\cdot\hat{\textbf{s}}_{j}), \end{equation}
where index $s$ or $s'$ refers to the subsystem of spin-up or spin-down ions.
Moreover, $s$ and $s'$ refer to different subsystems.
It leads to the polarization definition in the antiferromagnetic multiferroics
\begin{equation}\label{MFMafmUEI P def AFM} \textbf{P}(\textbf{r},t)=
\int \Psi^{\dagger}(R,t)\sum_{i\in A\cup B}\delta(\textbf{r}-\textbf{r}_{i})
(\hat{\textbf{d}}_{i}\Psi(R,t))dR, \end{equation}
where $s=A$ refers to the subsystem of spin-up ions,
and $s=B$ refers to the subsystem of spin-down ions.
The summation in equation (\ref{MFMafmUEI P def AFM}) explicitly specifies
that index $s$ in operator (\ref{MFMafmUEI edm operator Mod s}) belongs to both subspecies $A$ and $B$.
The same is true for the index $s'$, but they cannot belong to the same subspecies.
Definition (\ref{MFMafmUEI edm operator Mod s}) can be splitted on two partial polarizations
$\textbf{P}=\textbf{P}_{A}+\textbf{P}_{B}$
\begin{equation}\label{MFMafmUEI P def AFM A} \textbf{P}_{A}(\textbf{r},t)=
\int \Psi^{\dagger}(R,t)\sum_{i\in A}\delta(\textbf{r}-\textbf{r}_{i})
(\hat{\textbf{d}}_{i}\Psi(R,t))dR, \end{equation}
with
$\hat{d}_{i\in A}^{\alpha}=\sum_{j\in B}
\Pi_{ij}^{\alpha}(r_{ij})(\hat{\textbf{s}}_{i}\cdot\hat{\textbf{s}}_{j})$,
and
\begin{equation}\label{MFMafmUEI P def AFM B} \textbf{P}_{B}(\textbf{r},t)=
\int \Psi^{\dagger}(R,t)\sum_{i\in B}\delta(\textbf{r}-\textbf{r}_{i})
(\hat{\textbf{d}}_{i}\Psi(R,t))dR, \end{equation}
with
$\hat{d}_{i\in B}^{\alpha}=\sum_{j\in A}
\Pi_{ij}^{\alpha}(r_{ij})(\hat{\textbf{s}}_{i}\cdot\hat{\textbf{s}}_{j})$.
In order to make the splitting given by equations
(\ref{MFMafmUEI P def AFM A}) and (\ref{MFMafmUEI P def AFM B}),
we specify the subspecies $s=A$ or $s=B$ in operator (\ref{MFMafmUEI edm operator Mod s}).

We can calculate $\textbf{P}_{A}$ and $\textbf{P}_{B}$ separately
$$P^{\alpha}_{A}=
g_{0\Pi,AB}^{\alpha}\textbf{S}_{A}\textbf{S}_{B}
+\frac{1}{3}\frac{1}{2^{3}}g_{\Pi}^{\alpha}\biggl[
\triangle (\textbf{S}_{A}\textbf{S}_{B})$$
$$+2 \partial_{\mu} [S^{\nu}_{A} \partial_{\mu} S^{\nu}_{B} -S^{\nu}_{B} \partial_{\mu} S^{\nu}_{A}]$$
\begin{equation}\label{MFMafmUEI P def expanded AFR A}
+S^{\nu}_{A}(\triangle S^{\nu}_{B})+S^{\nu}_{B}(\triangle S^{\nu}_{A})
-2(\partial_{\mu} S^{\nu}_{A})(\partial_{\mu} S^{\nu}_{B})
\biggr],
\end{equation}
and
$$P^{\alpha}_{B}=
g_{0\Pi,AB}^{\alpha}\textbf{S}_{A}\textbf{S}_{B}
+\frac{1}{3}\frac{1}{2^{3}}g_{\Pi}^{\alpha}\biggl[
\triangle (\textbf{S}_{A}\textbf{S}_{B})$$
$$+2 \partial_{\mu} [S^{\nu}_{B} \partial_{\mu} S^{\nu}_{A} -S^{\nu}_{A} \partial_{\mu} S^{\nu}_{B}]$$
\begin{equation}\label{MFMafmUEI P def expanded AFR B}
+S^{\nu}_{A} \triangle S^{\nu}_{B} +S^{\nu}_{B} \triangle S^{\nu}_{A}
-2(\partial_{\mu} S^{\nu}_{A})(\partial_{\mu} S^{\nu}_{B})
\biggr].
\end{equation}
Equations (\ref{MFMafmUEI P def expanded AFR A}) and (\ref{MFMafmUEI P def expanded AFR B}) have similar structure.
The difference between them is in the replacement of subindexes $A\leftrightarrow B$.
We obtain the major contribution appearing from the first order of the expansion on the relative distance.
It is indicated by the coefficient $g_{0\Pi,AB}^{\alpha}$.
We also find the correction to the major term.
These corrections contain the coordinate derivatives of the spin density.
These terms are indicated by the coefficient $g_{\Pi,AB}^{\alpha}$.
The definitions of $g_{0\Pi,AB}^{\alpha}$ and $g_{\Pi,AB}^{\alpha}$ are identical
to the definitions of $g_{0\Pi}^{\alpha}$ and $g_{\Pi}^{\alpha}$ presented after equation (\ref{MFMafmUEI P def expanded FR}).

Next, we combine equations
(\ref{MFMafmUEI P def expanded AFR A}) and (\ref{MFMafmUEI P def expanded AFR B})
to get an expression for polarization of the whole system
\begin{equation}\label{MFMafmUEI P def expanded AFR}
P^{\alpha}=
2g_{0\Pi,AB}^{\alpha}\textbf{S}_{A}\textbf{S}_{B}
+\frac{1}{6}g_{\Pi,AB}^{\alpha}
(S^{\nu}_{A} \triangle S^{\nu}_{B} +S^{\nu}_{B} \triangle S^{\nu}_{A}).
\end{equation}
Equation (\ref{MFMafmUEI P def expanded AFR}) shows
that the direction of polarization is not related to the direction of the spin or the direction of change of the spin density in space.
The direction is defined by the coefficients $g_{0\Pi,AB}^{\alpha}$ and $g_{\Pi,AB}^{\alpha}$.
Below, we show that both constants have the same direction.
It is the direction of the shift of the nonmagnetic ion from the line of the magnetic ions.

We also represent the major term in the polarization of
via vectors $\textbf{L}$ and $\mbox{\boldmath $\Sigma$}$:
\begin{equation}\label{MFMafmUEI P def expanded AFR L M}
P^{\alpha}=
\frac{1}{2}g_{0\Pi,AB}^{\alpha}(\mbox{\boldmath $\Sigma$}^{2}-\textbf{L}^{2})
\approx -\frac{1}{2}g_{0\Pi,AB}^{\alpha}\textbf{L}^{2}.
\end{equation}

\section{Derivation of spin current model}

\subsection{The spin current model as the consequence of the momentum balance equation with spin-orbit interaction}

It is possible to use the electric dipole moment operator (\ref{MFMafmUEI edm operator Mod})
for the further derivation of the macroscopic polarization evolution equation
as an addition to the
Landau--Lifshitz--Gilbert equation
for the study of perturbations and structures in the multiferroics.
However, we are going to derive and justify the electric dipole moment operator (\ref{MFMafmUEI edm operator Mod})
starting from the quantum microscopic theory.
Therefore, we consider the many-particle Pauli equation (\ref{MFMafmUEI Sch eq})
with the following Hamiltonian
$$\hat{H}
=\sum_{i=1}^{N}\biggl[-\hat{\textbf{d}}_{i}\cdot\textbf{E}_{i}
-\hat{\mbox{\boldmath $\mu$}}_{i}\cdot\textbf{B}_{i}
-\frac{1}{2mc}(\hat{\mbox{\boldmath $\mu$}}_{i}\cdot[\textbf{E}_{i}\times\hat{\textbf{p}}_{i}])$$
\begin{equation}\label{MFMafmUEI Ham}
-\frac{1}{2}\sum_{j=1, j\neq i}^{N} \biggl(U_{ij}
\hat{\textbf{s}}_{i}\cdot\hat{\textbf{s}}_{j}
+\textbf{D}_{ij}\cdot
[\hat{\textbf{s}}_{i}\times\hat{\textbf{s}}_{j}] \biggr)\biggr],\end{equation}
where
$m$ is the mass of atom/ion,
$c$ is the speed of light in the vacuum,
$\hat{\textbf{d}}_{i}$ is the electric dipole moment operator,
being defined via the displacement of ions with different charges,
its relation to the spins of ions will be found below,
$\textbf{E}_{i}$ is the electric field,
acting on the $i$-th dipole,
$U_{ij}=U(\textbf{r}_{i}-\textbf{r}_{j})$
is the exchange integral of the Heisenberg Hamiltonian as the function of the interparticle distance,
$\textbf{D}_{ij}$ is the Dzylaoshinskii vector constant.
The Dzylaoshinskii vector
has the structure related to the relative position of two magnetic ions and one nonmagnetic ion (the ligand ion)
\cite{Pyatakov UFN 12}, \cite{Khomskii JETP 21}.
It can be presented via the vector product of the radius-vectors of magnetic ions relatively nonmagnetic ion \cite{Pyatakov UFN 12}.
Overwise, it can be presented as the vector product of the relative position of two magnetic ions
and the shift of the nonmagnetic ion from the line connecting two magnetic ions \cite{Khomskii JETP 21}
$\textbf{D}_{ij}\sim\textbf{r}_{ij}\times\mbox{\boldmath $\delta$}$.
This simple formula is useful,
if we discuss one cell.
However, if we consider the whole crystal macroscopically,
we need to specify that we consider two neighboring ions.
Hence, we introduce a function,
which decreases (drops to zero) at the distances beyond the period of the crystal cell.
So, we have the following structure
$\textbf{D}_{ij}=\beta(r_{ij})\textbf{r}_{ij}\times\mbox{\boldmath $\delta$}$.

The single-ion anisotropy is considered in Ref. \cite{Tokura RPP 14} (see p. 34, equations (39), (40)),
but we do not include it in our model at this stage.
Equation (41) of Ref. \cite{Tokura RPP 14} also presents the biquadratic interaction
(see also the model in Ref. \cite{Mochizuki PRL 10}),
which is partially considered within our model for the spin evolution equation
\cite{Andreev Trukhanova arXiv 23 05}.

Let us also describe the physical meaning of terms in the Hamiltonian (\ref{MFMafmUEI Ham}).
The first term is the action of the electric field on the electric dipole moment.
The second term is the action of the magnetic field on the magnetic moment.
The third term is the spin-orbit interaction
showing the action of the electric field on the moving magnetic moment.
The fourth term is the Coulomb exchange interaction presented by the Heisenberg Hamiltonian.
The last term is the Dzylaoshinskii-Moriya interaction.

It is essential to specify
that the electric dipole moment is related to the group of magnetic and nonmagnetic ions.
However, in Sec. III,
we contracted the electric dipole moment operator associated with each magnetic ion.
This meaning of the operator is implicated in the Hamiltonian.

\subsection{The spin current model in ferromagnetic materials}

Here, we derive the macroscopic polarization corresponding to the dipole moment operator (\ref{MFMafmUEI edm operator Mod})
via the spin current model.
Here, we also show that the spin current model follows from the momentum balance equation.
Therefore, we derive the momentum balance equation corresponding to the Hamiltonian (\ref{MFMafmUEI Ham}).

To derive the momentum balance equation,
we need to define the momentum density via the many-particle wave function
as the quantum average of the momentum operator of each particle
\begin{equation}\label{MFMafmUEI p def} \textbf{p}(\textbf{r},t)=
\frac{1}{2}\int \sum_{i}\delta(\textbf{r}-\textbf{r}_{i})\biggl(\Psi_{S}^{\dagger}(R,t)
\hat{\textbf{p}}_{i}\Psi_{S}(R,t)+h.c.\biggr)dR ,\end{equation}
where $h.c.$ is the Hermitian conjugation,
and
$\textbf{p}_{i}=-\imath\hbar\nabla_{i}$ is the momentum operator of $i$-th particle.

We consider the time derivative of the momentum density (\ref{MFMafmUEI p def}).
The time derivative acts on the wave function,
while the time derivative of the wave function is replaced with Hamiltonian in accordance with the Pauli equation
$$ \partial_{t}\textbf{p}(\textbf{r},t)=
\frac{1}{2}\frac{\imath}{\hbar}\int \sum_{i}\delta(\textbf{r}-\textbf{r}_{i})\times$$
\begin{equation}\label{MFMafmUEI p time der via H}\times\biggl(\Psi^{\dagger}(R,t)
[\hat{H},\hat{\textbf{p}}_{i}]\Psi(R,t)+h.c.\biggr)dR. \end{equation}
Further calculation depends on the explicit form of the Hamiltonian.
Moreover, different interaction terms are considered in different approximations.
The first three terms in the Hamiltonian (\ref{MFMafmUEI Ham}) can be considered straightforwardly.
However, two last terms can be considered with the account of the short-range nature of these interactions.
It leads to the following momentum balance equation
$$\partial_{t}\textbf{p}=P^{\beta}\nabla E^{\beta}+\gamma S^{\beta}\nabla B^{\beta}
+\frac{\gamma}{2mc}\varepsilon^{\beta\gamma\delta}J^{\delta\gamma}(\nabla E^{\beta})$$
\begin{equation}\label{MFMafmUEI p evolution long H}
+g_{0u} S^{\beta}\nabla S^{\beta}
+\textbf{F}_{DM},\end{equation}
where
$g_{0u}=\int U(r)d\textbf{r}$,
$J^{\delta\gamma}$ is the spin-current tensor,
$P^{\beta}$ is the polarization or the electric dipole moment density (\ref{MFMafmUEI P def}),
$\textbf{F}_{DM}$ is the force density of the Dzylaoshinskii-Moriya interaction:
\begin{equation}\label{MFMafmUEI F by DM FM}
\textbf{F}_{DM}=\frac{1}{3}g_{(\beta)}
\biggl((\mbox{\boldmath $\delta$}\cdot\textbf{S})\nabla(\nabla\cdot \textbf{S})
-(\textbf{S}\cdot\nabla)\nabla(\mbox{\boldmath $\delta$}\cdot\textbf{S})\biggr),
\end{equation}
with
$g_{(\beta)}=\int \xi^{2}\beta(\xi)d\mbox{\boldmath $\xi$}$.
So, physical meaning of terms in equation is the same like in Hamiltonian (\ref{MFMafmUEI Ham}),
they are also placed in the same order.

Let us consider the stationary regime,
where the momentum density does not change in time $\partial_{t}\textbf{p}=0$.
Therefore, the right-hand side of the momentum balance equation (\ref{MFMafmUEI p evolution long H}) should be equal to zero.
The balance of the second and fourth terms gives an equilibrium magnetic field formed by magnetic moments due to the exchange interaction
$\textbf{B}=-g_{0u} \textbf{S}/\gamma$.
The last term can be equal to zero in equilibrium
if the spin polarization is perpendicular to the shift of the ligand ion
from the line connecting neighboring magnetic ions
$(\mbox{\boldmath $\delta$}\cdot\textbf{S})=0$
or it can be equal to zero at the nontrivial balance of two terms in equation (\ref{MFMafmUEI F by DM FM}).
However, our goal is to obtain the spin current model,
which follows from the balance of the first and third terms.
It can happen in the arbitrary electric field
if polarization is balanced by the spin current appearing in the spin-orbit interaction
\begin{equation}\label{MFMafmUEI SC model} P^{\mu}
=\frac{\gamma}{2mc}\varepsilon^{\mu\alpha\beta}J^{\alpha\beta}.\end{equation}

We obtain the spin-current model of the polarization with no particular relation to the form of the spin-current.
Hence, equation (\ref{MFMafmUEI SC model}) can be applied for the derivation of the polarization caused by different mechanisms.
The mean-field part of the spin-orbit interaction plays main role in the derivation of the spin current model, i.e. in the connection between the polarization and the antisymmentric part of the spin-current tensor.
To the best of our knowledge we are not familiar
with the contribution of the symmetric part of the spin current tensor in the polarization of multiferroics.

\subsection{The spin current model in antiferromagnetic materials}

The first three terms in the Hamiltonian (\ref{MFMafmUEI Ham})
and the momentum balance equation (\ref{MFMafmUEI p evolution long H})
have the same form for the ferromagnetic and antiferromagnetic materials.
A difference appears for the Coulomb exchange interaction and the Dzylaoshinskii-Moriya interaction.
The force fields for these interactions are obtained in the following forms
\begin{equation}\label{MFMafmUEI F HH antiF}
\textbf{F}_{HH,s}=g_{0u,ss'}S_{s}^{\beta}\nabla S_{s'}^{\beta},
\end{equation}
and
$$\textbf{F}_{DM,s}=\frac{1}{3}g_{(\beta),ss'}
\biggl((\mbox{\boldmath $\delta$}\cdot\textbf{S}_{s})\nabla(\nabla\cdot \textbf{S}_{s'\neq s}) $$
\begin{equation}\label{MFMafmUEI F DM antiF}
-(\textbf{S}_{s}\cdot\nabla)\nabla(\mbox{\boldmath $\delta$}\cdot\textbf{S}_{s'\neq s})\biggr). \end{equation}
Hence, we derive the momentum balance equation for each subspecies
$$\partial_{t}\textbf{p}_{s}=P^{\beta}_{s}\nabla E^{\beta}+\gamma S^{\beta}_{s}\nabla B^{\beta}$$
\begin{equation}\label{MFMafmUEI p evolution long H sub}
+\frac{\gamma}{2mc}\varepsilon^{\beta\gamma\delta}J^{\delta\gamma}_{s}(\nabla E^{\beta})
+\textbf{F}_{HH,s}
+\textbf{F}_{DM,s}.\end{equation}
The electric dipole moment is related to the group of ions,
but the operator definition
is recontracted to the form of operator associated with each magnetic ion.
Hence, we get the partial polarization of each subspecies from the momentum balance equation of each subspecies.
It gives us the representation of the partial polarization via the partial spin current
$P^{\mu}_{s}
=\frac{\gamma}{2mc}\varepsilon^{\mu\alpha\beta}J^{\alpha\beta}_{s}$.
Complete polarization of the sample is the combination of partial polarizations
$P^{\mu}=P^{\mu}_{A}+P^{\mu}_{B}
=\frac{\gamma}{2mc}\varepsilon^{\mu\alpha\beta}(J^{\alpha\beta}_{A}+J^{\alpha\beta}_{B})$.
Therefore, equation (\ref{MFMafmUEI SC model}) is reestablished for the antiferromagnetic materials.

\subsection{Dzylaoshinskii-Moriya spin current and related polarization in ferromagnetic materials}

We presented the derivation of the spin current model.
It appears due to the relativistic spin-orbit interaction.
Its further application requires an example of the spin current related to a specific physical mechanism.
The explicit form of the spin-current tensor can be found in the spin evolution equation
(Landau--Lifshitz--Gilbert equation).
In this work, we focus on the justification of operator (\ref{MFMafmUEI edm operator Mod}),
which contains the vector coefficient of proportionality
between the combination of the spin operators and the electric dipole moment operator.
In Hamiltonian (\ref{MFMafmUEI Ham})
we have two interactions containing inexplicitly defined space dependencies.
They are the two last terms corresponding to the exchange Coulomb interaction given by the Heisenberg Hamiltonian
and the Dzylaoshinskii-Moriya interaction, correspondingly.
The Heisenberg Hamiltonian contains a scalar function,
so we expect that it is not related to the considering regime.
In contrast, the Dzylaoshinskii-Moriya interaction is proportional to the vector Dzylaoshinskii constant,
so it can give a mechanism for the vector constant appearance in operator (\ref{MFMafmUEI edm operator Mod}).
To check the described suggestion,
we need to consider the Dzylaoshinskii-Moriya interaction contribution in the spin evolution equation.
Similarly to Sec. II,
we use the quantum hydrodynamic method and include the short-range nature of the Dzylaoshinskii-Moriya interaction.
As the result, we find the partial contribution of the Dzylaoshinskii-Moriya interaction in the spin evolution equation
\begin{equation}\label{MFMafmUEI s evolution long H}
\partial_{t}\textbf{S}_{s}=\textbf{T}_{DM}, \end{equation}
where
\begin{equation}\label{MFMafmUEI spin torque DM F}
\textbf{T}_{DM}=\frac{1}{3}g_{(\beta)}
\biggl((\textbf{S}_{s}\cdot[\mbox{\boldmath $\delta$}\times\nabla])\textbf{S}_{s}
-S_{s}^{\beta}[\mbox{\boldmath $\delta$}\times\nabla]S_{s}^{\beta}\biggr) ,\end{equation}
with $g_{(\beta)}=\int \xi^{2}\beta(\xi)d\mbox{\boldmath $\xi$}$.

Ref. \cite{Wang PRL 15}
is based on the total free energy with a nonexplicit account of the Dzylaoshinskii-Moriya interaction.
Its explicit contribution is given in the Landau-Lifshitz-Gilbert equation \cite{Tatara PR 08} in the form of the torque
$\textbf{T}=\gamma \tilde{D}\textbf{M}\times(\nabla\times \textbf{M})$
(see eq. 2 and text after eq. 2 of Ref. \cite{Wang PRL 15}).
The expression we obtain in this paper (\ref{MFMafmUEI spin torque DM F}) shows a similar structure
at the replacement $g_{(\beta)}[\mbox{\boldmath $\delta$}\times\nabla]$ on  $\tilde{D}\nabla$.
The difference partially appears due to the account of structure of
the Dzylaoshinskii vector constant $\textbf{D}_{ij}=\beta(r_{ij})\textbf{r}_{ij}\times\mbox{\boldmath $\delta$}$.
These expressions give different directions of the spin torque vector.

The spin wave is considered described in Ref. \cite{Wang PRL 15} as a small fluctuation of the static domain-wall profile.
As a limit, a simple dispersion dependence is found for location away from the domain-wall center,
where the magnetization is uniform in the domains (see eq. 9 of Ref. \cite{Wang PRL 15}).
The dispersion relations appears to be asymmetric
outside the domain-wall in accordance with works \cite{Zakeri PRL 10}, \cite{Moon PRB 13}.
Let us specify that in Ref. \cite{Zakeri PRL 10} experimentally demonstrated
(on an Fe double layer grown on W(110))
the Dzyaloshinskii-Moriya leads to an asymmetric spin-wave dispersion relation.

Ref. \cite{Moon PRB 13} presents a translation of the Dzyaloshinskii-Moriya interaction Hamiltonian
(like the last term in eq. (\ref{MFMafmUEI Ham}))
to a continuum
model of the energy density with magnetization direction  and symmetry breaking in the $y$-direction.

In order to compare our model with eq. (2) of Ref. \cite{Moon PRB 13},
we calculate the correlationless limit of the energy density.
The definition of the energy density is
$$E_{DM}(\textbf{r},t)=
-\frac{1}{2}\int \sum_{i,j=1, j\neq i}^{N}\delta(\textbf{r}-\textbf{r}_{i})\times$$
\begin{equation}\label{MFMafmUEI Energy DM def}
\times \Psi_{S}^{\dagger}(R,t)
\textbf{D}_{ij}\cdot [\hat{\textbf{s}}_{i}\times\hat{\textbf{s}}_{j}]\Psi_{S}(R,t)dR .\end{equation}
It gives us the following approximate expression
\begin{equation}\label{MFMafmUEI Energy DM expansion}
E_{DM} \approx \frac{1}{6}g_{(\beta)}\delta^{\alpha}
(S^{\alpha}(\nabla\cdot \textbf{S})-(\textbf{S}\cdot\nabla)S^{\alpha}),
\end{equation}
where we use $\textbf{D}_{ij}=\beta(r_{ij})\textbf{r}_{ij}\times\mbox{\boldmath $\delta$}$.
Let us repeat here the eq. (2) Ref. \cite{Moon PRB 13} for the comparison
$E_{DM}=-D [ (\textbf{m}\times\partial_{x}\textbf{m})_{z}-(\textbf{m}\times\partial_{z}\textbf{m})_{x}]$,
where $\textbf{m}$ is the magnetization direction,
$(\textbf{b})_{x}$ is the $x$-projection of vector $\textbf{b}$,
and the $y$-direction is chosen in Ref. \cite{Moon PRB 13} as the anisotropy direction.
For the fixed concentration we have $\textbf{S}\sim \textbf{m}$
(otherwise the change of concentration would give contribution in the derivatives of the spin density $\textbf{S}$).
We see rather different vector structures in these expressions.
To continue our comparison,
we would assume $\textbf{D}_{ij}=\tilde{\beta}(r_{ij})\textbf{r}_{ij}$.
It leads to $E_{DM}=-(1/6)g_{(\tilde{\beta})}\textbf{S}\cdot[\nabla\times\textbf{S}]$,
but it also differs from even if we assume that there is no dependence on the chosen direction $y$.
The described difference can be caused by the different form of structure of the Dzyaloshinskii constant $\textbf{D}_{ij}$.
Our expression (\ref{MFMafmUEI Energy DM expansion}) corresponds to eq. 2 in Ref. \cite{Thiaville EPL 12},
up to the details of the coefficient.
Here, we presented the derivation for the arbitrary three-dimensional sample,
while the ultrathin films with perpendicular
easy axis, grown on a substrate with a capping
from the different material so that the structural inversion
symmetry is broken along the film normal, are considered in Ref. \cite{Thiaville EPL 12}.

The Dzylaoshinskii constant has a known particular numerical value for materials.
In contrast, constants like $g_{(\beta)}$ are novel parameters.
Hereby, we need to give some comment on the estimation of these parameters for better comparison with other works.
To get an approximate expression for $g_{(\beta)}$
if we consider explicit form for the function $\beta(r)$.
We choose a model expression for the function $\beta(r)$:
$\beta(r)=(D_{0}/(4\pi a \delta))\theta(r-a)$,
where $a$ is the distance between magnetic ions,
$\delta$ is the module of the vector of the ligand shift.
It leads to
$g_{(\beta)}=D_{0}a^{4}/5\delta$.
Similar analysis for the constant like $g_{0u}$ is given in Ref. \cite{Andreev Trukhanova Vestn 23}.

Equation (\ref{MFMafmUEI spin torque DM F}) shows
that the spin-torque caused by the Dzylaoshinskii-Moriya interaction
cannot be represented as the divergence of the spin-current tensor.
However, the second term in (\ref{MFMafmUEI spin torque DM F}) can be represented in the required form
$S_{s}^{\delta}\varepsilon^{\alpha\beta\gamma}\delta^{\beta}\nabla^{\gamma}S_{s}^{\delta}$
$=\nabla^{\gamma}[(1/2)\varepsilon^{\alpha\beta\gamma}\delta^{\beta}\textbf{S}_{s}^{2}]$.
We can use this part for the calculation of the polarization.
However, the first term in (\ref{MFMafmUEI spin torque DM F}) gives some freedom in the interpretation of the spin current
since it allows to get an additional term.

Let us start the analysis of the polarization using the simplest form of the Dzylaoshinskii-Moriya spin current
$\textbf{T}_{DM}=-\partial_{\beta}J^{\alpha\beta}_{DM}$
\begin{equation}\label{MFMafmUEI SC DM 1} J^{\alpha\beta}_{DM}=-\frac{1}{6}g_{(\beta)}
\varepsilon^{\alpha\beta\gamma}\delta^{\gamma}\textbf{S}_{s}^{2}. \end{equation}
It leads to the following form of the macroscopic polarization
\begin{equation}\label{MFMafmUEI P DM from SCm 1}
P^{\mu}_{DM}=\frac{\gamma}{2mc}\varepsilon^{\mu\alpha\beta}J^{\alpha\beta}_{DM}
=-\frac{1}{6}\frac{\gamma}{mc}g_{(\beta)}
\delta^{\mu}\textbf{S}_{s}^{2},\end{equation}
which corresponds to the first term on the right-hand side of equation (\ref{MFMafmUEI P def expanded FR}),
which is derived from operator (\ref{MFMafmUEI edm def i i+1}).
Let us to point out that
the collinear spin structure is formed by the Heisenberg exchange interaction,
while further formation of spin related electric polarization is formed
by the combination of the spin-orbit and the Dzyaloshinskii-Moriya interactions.

As a matter of discussion,
let us represent the spin torque (\ref{MFMafmUEI spin torque DM F}) in a form,
where the spin current is extracted from the first term in addition to the spin current following from the second term
$$\textbf{T}_{DM}=-\frac{1}{3}g_{(\beta)}
\biggl(
(\textbf{S}_{s}(\mbox{\boldmath $\delta$}\cdot[\nabla\times\textbf{S}_{s}])$$
\begin{equation}\label{MFMafmUEI spin torque DM F via SC}
+\nabla^{\beta}\{[\mbox{\boldmath $\delta$}\times\textbf{S}_{s}]^{\beta} \textbf{S}_{s}\}
+\frac{1}{2}[\mbox{\boldmath $\delta$}\times\nabla]\textbf{S}_{s}^{2}\biggr)
.\end{equation}
It shows that we can choose the spin current in the following alternative form
\begin{equation}\label{MFMafmUEI SC DM 2}
\tilde{J}^{\alpha\beta}_{DM}=
\frac{1}{3}g_{(\beta)} \delta^{\gamma}
\biggl[
S^{\alpha}_{s}\varepsilon^{\beta\gamma\delta} S_{s}^{\delta}
-\frac{1}{2}\varepsilon^{\alpha\beta\gamma} \textbf{S}_{s}^{2}
\biggr].
\end{equation}
This extended spin current leads to the following form of polarization
\begin{equation}\label{MFMafmUEI P DM from SCm 02} \tilde{P}^{\mu}_{DM}
=\frac{\gamma}{2mc}\varepsilon^{\mu\alpha\beta}\tilde{J}^{\alpha\beta}_{DM}
=-\frac{1}{6}\frac{\gamma}{mc}g_{(\beta)}
(\mbox{\boldmath $\delta$}\cdot\textbf{S}_{s})S_{s}^{\mu}.
\end{equation}
This form of the polarization completely differs from the structure following from the well-known electric dipole moment of the pair of ions
(\ref{MFMafmUEI edm def i i+1}) and (\ref{MFMafmUEI P def expanded FR}).
We want to mention that the spin torque following from the Heisenberg exchange interaction
appears as the divergence of the spin current tensor with no additional terms.

In the comparison of expressions (\ref{MFMafmUEI P def expanded FR}) and (\ref{MFMafmUEI P DM from SCm 1}),
we see that they are found in different approximations.
Polarization (\ref{MFMafmUEI P DM from SCm 1}) is obtained in the main order of the expansion,
while polarization (\ref{MFMafmUEI P def expanded FR}) contains corrections related to the second space derivative of the spin density.
To complete our comparison,
we can find corrections to the spin-torque caused by the Dzylaoshinskii-Moriya interaction (\ref{MFMafmUEI spin torque DM F})
in the next order of expansion.
Our calculation gives the following spin-torque
\begin{equation}\label{MFMafmUEI spin torque DM F3}
T^{\mu}_{DM,3}=
\frac{1}{30}g_{2(\beta)}\biggl[(\textbf{S}_{s}\cdot[\mbox{\boldmath $\delta$}\times\nabla])\triangle\textbf{S}_{s}
-S_{s}^{\beta}[\mbox{\boldmath $\delta$}\times\nabla]\triangle S_{s}^{\beta}
\biggr] ,\end{equation}
with $g_{2(\beta)}=\int \xi^{4}\beta(\xi)d\mbox{\boldmath $\xi$}$.

We need to extract a part of the spin torque,
which appears as the divergence of the spin current
$\partial^{\beta}J^{\alpha\beta}$.
In equation (\ref{MFMafmUEI spin torque DM F}) we used the term containing the scalar product of the spin densities
$S_{s}^{\beta}[\mbox{\boldmath $\delta$}\times\nabla]S_{s}^{\beta}$
(the second term on the right-hand side).
In equation (\ref{MFMafmUEI spin torque DM F3}) we follow the same approach and consider the term containing the scalar product of the spin densities
$S_{s}^{\beta}[\mbox{\boldmath $\delta$}\times\nabla]\triangle S_{s}^{\beta}$
(the second term on the right-hand side).
However, the term under consideration does not appear as the divergence of the second rank tensor.
We need to split it into two parts
$S_{s}^{\beta}[\mbox{\boldmath $\delta$}\times\nabla]\triangle S_{s}^{\beta}$
$=[\mbox{\boldmath $\delta$}\times\nabla](S_{s}^{\beta}\triangle S_{s}^{\beta})$
$-(\triangle S_{s}^{\beta})[\mbox{\boldmath $\delta$}\times\nabla]S_{s}^{\beta}$
and use the first of them to get the effective spin current
\begin{equation}\label{MFMafmUEI SC DM 3} J^{\alpha\beta}_{DM3}=-\frac{1}{5}\frac{1}{6}g_{2(\beta)}
\varepsilon^{\alpha\beta\gamma}\delta^{\gamma}(\textbf{S}_{s}\cdot\triangle\textbf{S}_{s}). \end{equation}
It leads to the following polarization
\begin{equation}\label{MFMafmUEI P DM from SCm 3}
P^{\mu}_{DM}=\frac{\gamma}{2mc}\varepsilon^{\mu\alpha\beta}J^{\alpha\beta}_{DM}
=-\frac{1}{6}\frac{1}{5}\frac{\gamma}{mc}g_{2(\beta)}
\delta^{\mu}(\textbf{S}_{s}\cdot\triangle\textbf{S}_{s}).\end{equation}
We see the spin structure in polarization (\ref{MFMafmUEI P DM from SCm 3})
corresponds to the spin structure in the second term on the right-hand side in equation
(\ref{MFMafmUEI P def expanded FR}).
Hence, we justify equation (\ref{MFMafmUEI P def expanded FR}) found from the electric dipole moment (\ref{MFMafmUEI edm def i i+1})
using the spin current model with the spin-current related to the Dzylaoshinskii-Moriya interaction.
Moreover, our calculations in two major orders of expansion give the interpretation of the direction of the vector constant $\Pi^{\alpha}(\xi)$.
We see that it is parallel to the shift of the ligand ion from the line connecting neighboring
magnetic ions $\delta^{\alpha}$.
The complite polarization is also parallel to this direction.
It also corresponds to the microscopic meaning of the electric dipole moment as a shift of ions of opposite charges.

In Sec. III, we found the macroscopic polarization (\ref{MFMafmUEI P def expanded FR})
corresponding to the electric dipole moment (\ref{MFMafmUEI edm def i i+1}).
Here, we found macroscopic polarization
using the momentum balance equation with the spin-orbit interaction
and the spin-current caused by the Dzylaoshinskii-Moriya interaction.
These expressions have the same structure.
It allows us to give a physical interpretation of the vector constant in the electric dipole moment (\ref{MFMafmUEI edm def i i+1}).
Let us compare the polarization given by the first term in equation (\ref{MFMafmUEI P def expanded FR})
with equation (\ref{MFMafmUEI P DM from SCm 1}).
It gives the following relation
$g_{0\Pi}^{\alpha}\textbf{S}_{s}^{2}=-\frac{1}{6}\frac{\gamma}{mc}g_{(\beta)}
\delta^{\mu}\textbf{S}_{s}^{2}$,
where we can drop the square of the spin density $\textbf{S}_{s}^{2}$ and compare the coefficients.
Basically, we need to compare the functions under the integrals.
We have two options.
First, we equate the functions under integrands
and find
$\Pi^{\alpha}(\xi)=-(1/6)(\gamma/mc)\xi^{2}\beta(\xi)\delta^{\alpha}$.
Second, we transform the left-hand side by integration by parts,
so we obtain $g_{0\Pi}^{\alpha}=-(1/3)\int\xi(\partial\Pi^{\alpha}(\xi)/\partial\xi)d^{3}\xi$.
Next, we equate the functions under integrands
and obtain $\partial\Pi^{\alpha}(\xi)/\partial\xi=(\gamma/2mc)\xi\beta(\xi)\delta^{\alpha}$.

We found a relation between the empirically introduced function $\Pi^{\alpha}(\xi)$,
which is the coefficient of proportionality in the electric dipole moment (\ref{MFMafmUEI edm def i i+1})
and the function $\beta(\xi)$ appearing in the Dzylaoshinskii-Moriya interaction.
The Dzylaoshinskii-Moriya interaction is the exchange part of the spin-orbit interaction,
and the coefficient $\textbf{D}_{ij}$ is the exchange integral.
It is similar to the exchange integral in the Heisenberg exchange interaction,
where the exchange part of the Coulomb interaction is considered.

In order to prove the found expression for function $\Pi^{\alpha}(\xi)$
we considered the next order of expansion for the polarization definition
(\ref{MFMafmUEI P def expanded FR})
and the Dzylaoshinskii-Moriya spin-torque giving the effective spin-current
and corresponding polarization (\ref{MFMafmUEI P DM from SCm 3}).
Hence, we compare the second term on the right-hand side of equation
(\ref{MFMafmUEI P def expanded FR})
with polarization (\ref{MFMafmUEI P DM from SCm 3}).
Dropping equal spin structures,
we find
$\frac{1}{6}g_{\Pi}^{\alpha}
=-\frac{1}{6}\frac{1}{5}\frac{\gamma}{mc}g_{2(\beta)}
\delta^{\mu}$.
Here we need to compare the functions under integral and prove the relation
between $\partial\Pi^{\alpha}(\xi)/\partial\xi$ and $\beta(\xi)$
obtained above.
We transform the left-hand side by integration by parts,
so we obtain $g_{\Pi}^{\alpha}=-(1/5)\int\xi^{3}(\partial\Pi^{\alpha}(\xi)/\partial\xi)d^{3}\xi$.
It leads to relation
\begin{equation}\label{MFMafmUEI Pi via beta}
\frac{\partial\Pi^{\alpha}(\xi)}{\partial\xi}=\frac{\gamma}{2mc}\xi\beta(\xi)\delta^{\alpha}, \end{equation}
which is presented above from the first order of expansion.

The spin torque caused by the Heisenberg exchange interaction can be presented
as the divergence of the corresponding spin current tensor \cite{Landau vol 9}
$g_{u}\varepsilon^{\alpha\beta\gamma}S_{s}^{\beta}\triangle S_{s}^{\gamma}
=\partial_{\delta}(g_{u}\varepsilon^{\alpha\beta\gamma}S_{s}^{\beta}\partial_{\delta} S_{s}^{\gamma})
=-\partial_{\delta}J^{\alpha\delta}_{HH}$.
This spin current tensor can be placed in the polarization obtained in the spin current model
$P^{\mu}
=\frac{\gamma}{2mc}\varepsilon^{\mu\alpha\beta}J^{\alpha\beta}$.
It gives polarization coinciding with the result of M. Mostovoy \cite{Mostovoy PRL 06}.
The method demonstrated in Sec. III can be applied to the operator
$\textbf{d}_{ij}= \alpha_{ij}[\textbf{r}_{ij}\times[\textbf{s}_{i}\times\textbf{s}_{j}]]$
in order to rederive the result of M. Mostovoy \cite{Mostovoy PRL 06}.
So, we can conclude that this result follows from the Heisenberg exchange interaction.
This comment is placed here for comparison with the results of our paper.

\subsection{Dzylaoshinskii-Moriya spin current and related polarization in antiferromagnetic materials}

In this section,
we need to consider the spin-torque caused by the Dzylaoshinskii-Moriya interaction.
Particularly, we need to consider the interaction between different subspecies in the antiferromagnetic samples.
Our calculations give the following form of the $s$ subspecies spin evolution equation
under the Dzylaoshinskii-Moriya interaction with $s'$ subspecies:
\begin{equation}\label{MFMafmUEI s evolution DM antiF}
\partial_{t}\textbf{S}_{s}=\textbf{T}_{DM,s'\neq s}, \end{equation}
where
\begin{equation}\label{MFMafmUEI spin torque DM antiF s}
\textbf{T}_{DM,s'\neq s}=\frac{1}{3}g_{(\beta)AB}
\biggl((\textbf{S}_{s}\cdot[\mbox{\boldmath $\delta$}\times\nabla])\textbf{S}_{s'\neq s}
-S_{s}^{\beta}[\mbox{\boldmath $\delta$}\times\nabla]S_{s'\neq s}^{\beta}\biggr) .\end{equation}
The general structure of the obtained spin torque is similar to the torque existing
under interaction of the ions of the same subspecies (\ref{MFMafmUEI spin torque DM F}).
However, there is an essential difference related to the appearance of two kinds of subindexes, $s$ and $s'$.
So, we find no term,
which can be rewritten as the spin current.

In order to solve the described problem,
we suggest the following step.
We need to consider the sum of spin torques $\textbf{T}_{DM,A}$ and $\textbf{T}_{DM,B}$
instead of the sum of partial spin currents:
$$\textbf{T}_{DM}=\frac{1}{3}g_{(\beta)AB}
\biggl((\textbf{S}_{A}\cdot[\mbox{\boldmath $\delta$}\times\nabla])\textbf{S}_{B}$$
\begin{equation}\label{MFMafmUEI spin torque DM antiF AB}
+(\textbf{S}_{B}\cdot[\mbox{\boldmath $\delta$}\times\nabla])\textbf{S}_{A}
-[\mbox{\boldmath $\delta$}\times\nabla](S_{A}^{\beta}S_{B}^{\beta})\biggr) .\end{equation}
So, a part of combined spin torque can be presented as a "combined" spin current.
So, we would be able to derive the spin polarization of the full system instead of partial polarizations.
Anyway, the partial polarizations are intermediate theoretical constructions,
which have no physical meaning since the polarization formation is related to magnetic ions of both subspecies
(and ions of the nonmagnetic subspecies).
Similarly to the last term in equation (\ref{MFMafmUEI spin torque DM F}),
we see that the last term in equation (\ref{MFMafmUEI spin torque DM antiF AB}) gives us the effective spin current
caused by the intersubspecies Dzylaoshinskii-Moriya interaction
\begin{equation}\label{MFMafmUEI SC DM antiF} J^{\alpha\beta}_{DM}=-\frac{1}{3}g_{(\beta)AB}
\varepsilon^{\alpha\beta\gamma}\delta^{\gamma}(\textbf{S}_{A}\cdot\textbf{S}_{B}). \end{equation}
It leads to the polarization of antiferromagnetic materials
\begin{equation}\label{MFMafmUEI P DM from SCm 1 AFM}
P^{\mu}_{DM}=\frac{\gamma}{2mc}\varepsilon^{\mu\alpha\beta}J^{\alpha\beta}_{DM}
=-\frac{1}{3}\frac{\gamma}{mc}g_{(\beta)AB}
\delta^{\mu}(\textbf{S}_{A}\cdot\textbf{S}_{B}).\end{equation}

\subsection{On the other types of mechanisms of the polarization formation}

In this paper we are focused on the contribution of the Dzylaoshinskii-Moriya spin current
in the spin current model in order to derive the symmetric form of the electric polarization in multiferroics.
However, this is one of three known forms of the polarization.
One of them is the antisymmetric form of the electric polarization.
Corresponding polarization can be obtained via the magnon spin current existing due to the Heisenberg exchange interaction
(it is briefly described in the final part of Sec. IV.D.).
The biquadrtic exchange can give a correction to this result.
The third type of the polarization is the spin dependent p-d hybridization mechanism.
At the current state of our research we cannot suggest any interaction which provides a suitable spin current.
Here we demonstrated that the spin current model can be applied for the symmetric form of the electric polarization
in addition to the well-known explanation of the antisymmetric form of polarization,
but it is not extended to the third mechanism of the polarization formation.

\section{Polarization evolution equation}

For the derivation of the polarization evolution equation,
we use the Hamiltonian (\ref{MFMafmUEI H Ham short HH and mB}),
where we include no relativistic interactions.
The polarization itself is caused by the relativistic effects,
hence, relativistic interactions
(the spin-orbit interaction, the Dzylaoshinskii-Moriya interaction,
and the evolution of the dipole moment under the action of the external electric field
due to the relativistic nature of the electric dipole moment)
give the relatively small effect.

In order to derive the polarization evolution equation,
we consider the definition of the polarization in terms of the microscopic many-particle wave function
(\ref{MFMafmUEI P def}),
with operator (\ref{MFMafmUEI edm operator Mod}).
We consider the time derivative of this definition.
The time derivative acts on the wave functions under the integral.
We find for Hamiltonian (\ref{MFMafmUEI H Ham short HH and mB}) the following intermediate form of the polarization evolution equation
\begin{equation}\label{MFMafmUEI P time der via H} \partial_{t}\textbf{P}(\textbf{r},t)=
\frac{\imath}{\hbar}\int \sum_{i}\delta(\textbf{r}-\textbf{r}_{i})\Psi^{\dagger}(R,t)
[\hat{H},\hat{\textbf{d}}_{i}]\Psi(R,t)dR. \end{equation}

The first term
contains dependence
on two particles $i$ and $j$ in the functions
$\delta(\textbf{r}-\textbf{r}_{i})$, $B_{i}^{\alpha}$ and $\Pi^{\alpha}_{ij}$
placed under the integral in equation (\ref{MFMafmUEI P time der via H}).
In the following calculations of this term,
we need to include the strong decrease of function $\Pi^{\alpha}_{ij}$
with the increase of the interparticle distance.
We need to make the transition to relative interparticle distance similarly to Sec. III,
where we made analysis of the definition of polarization (\ref{MFMafmUEI P def}).

The second term
has a more complex structure.
It depends on three particles $i$, $j$ and $k$ in the functions
placed under the integral in equation (\ref{MFMafmUEI P time der via H}).
Therefore, we need to introduce the center of mass and the relative distances for three particles.
In our calculations, we use the following substitution
$\textbf{r}_{i}=\textbf{R}_{ijn}+(2/3)\textbf{r}_{in}-(1/3)\textbf{r}_{jn}$,
$\textbf{r}_{j}=\textbf{R}_{ijn}-(1/3)\textbf{r}_{in}+(2/3)\textbf{r}_{jn}$,
and
$\textbf{r}_{n}=\textbf{R}_{ijn}-(1/3)\textbf{r}_{in}-(1/3)\textbf{r}_{jn}$,
where
$\textbf{R}_{ijn}=(\textbf{r}_{i}+\textbf{r}_{j}+\textbf{r}_{n})/3$,
$\textbf{r}_{in}\equiv\textbf{r}_{1}=\textbf{r}_{i}-\textbf{r}_{n}$,
$\textbf{r}_{jn}\equiv\textbf{r}_{2}=\textbf{r}_{j}-\textbf{r}_{n}$,
and
$\textbf{r}_{ij}\equiv\textbf{r}_{3}=\textbf{r}_{1}-\textbf{r}_{2}$.
It leads to the change of the element of volume in the configuration space
$dR=dR_{N-3}d\textbf{R}_{ijk}d\textbf{r}_{in}d\textbf{r}_{jn}$.
We use these substitutions
in the delta function
$\delta(\textbf{r}-\textbf{r}_{i})$
and in the many-particle wave function
$\Psi(R,t)=\Psi(...,\textbf{r}_{i}, ..., \textbf{r}_{j}, ..., \textbf{r}_{n}, ...,t)$.

After the described change of notations under the integral in equation (\ref{MFMafmUEI P time der via H})
we make an expansion on the relative distances.
It is possible due to the strong dependence of functions $\Pi^{\alpha}_{ij}$ and $U(r_{kj})$ on the relative distance.
Below, we present the results of our calculations for two regimes:
the ferromagnetic materials and the antiferromagnetic materials.

\subsection{Polarization evolution for the ferromagnetic materials}

In this subsection, we present the results of our derivation of the polarization evolution for the ferromagnetic materials
\begin{equation}\label{MFMafmUEI P time derivative FM}
\partial_{t}P^{\alpha}
=\frac{1}{3}\gamma g^{\alpha}_{\Pi}\varepsilon^{\beta\gamma\delta}
(\partial^{\mu}B^{\beta}) S^{\gamma}\nabla^{\mu}S^{\delta}
+G^{\alpha}(\textbf{S}\cdot [\nabla^{\mu}\textbf{S}\times \partial^{\mu}\triangle\textbf{S}]), \end{equation}
where
the first term is the contribution of the Zeeman energy (see the first term in Hamiltonian
(\ref{MFMafmUEI H Ham short HH and mB})),
it is obtained in this paper,
the last term is caused by the Coulomb exchange interaction
and obtained in \cite{Andreev 23 12}.
The following notations are used in equation (\ref{MFMafmUEI P time derivative FM}):
$\gamma$ is the gyromagnetic ratio,
$\hat{\mu}^{\alpha}_{i}=\gamma\hat{s}^{\alpha}_{i}$ is the magnetic moment,
and the vector interaction constant
\begin{equation}\label{MFMafmUEI G1 main}
G^{\alpha}=\frac{1}{3!}\frac{1}{3}g_{u}g_{\Pi}^{\alpha},\end{equation}
which is a combined interaction constant
with
$g_{u}=\int r^{2}U(r)d^{3}r$,
$g_{\Pi}^{\alpha}=\int r^{2}\Pi^{\alpha}(r)d^{3}r$.
Here we see one interaction constant related to the exchange integral
$g_{u}=\int r^{2}U(r)d^{3}r$, and
the interaction constant related to the function describing formation of the electric dipole moment
$g_{\Pi}^{\alpha}=\int r^{2}\Pi^{\alpha}(r)d^{3}r$.
The contribution of the Zeeman energy in the polarization evolution equation,
in the regime, where the electric dipole moment is proportional to the vector product of the spin operators,
is found in Ref. \cite{Andreev 23 11}.
It is related to another mechanism of the polarization formation in the multiferroic material.

\subsection{Polarization evolution for the antiferromagnetic materials}

In this subsection,
we consider the time evolution of polarization given by equation
(\ref{MFMafmUEI P def AFM A}),
which includes the structure of antiferromagnetic materials.
Here we have two subspecies $A$ and $B$,
so we derive the polarization evolution equation for each of them.
Let us present the result for one of the subspecies
$$\partial_{t}P^{\alpha}_{A}
=\frac{1}{6}\varepsilon^{\beta\gamma\delta}g_{\Pi}^{\alpha}\gamma (\partial^{\mu}B^{\gamma})
[S_{B}^{\delta}\partial^{\mu}S_{A}^{\beta}-S_{A}^{\beta}\partial^{\mu}S_{B}^{\delta}-\partial^{\mu}(S_{A}^{\beta}S_{B}^{\delta})]$$
$$+\frac{1}{6}\varepsilon^{\beta\gamma\delta}
[(g_{\Pi}^{\alpha}g_{0u}+2g_{0\Pi}^{\alpha}g_{u})S^{\beta}_{B}S^{\gamma}_{A}\triangle S_{A}^{\delta}$$
$$+(-g_{\Pi}^{\alpha}g_{0u}+2g_{0\Pi}^{\alpha}g_{u})S^{\beta}_{A}S^{\gamma}_{B}\triangle S_{B}^{\delta}$$
\begin{equation}\label{MFMafmUEI P time derivative AFM A}
+2g_{\Pi}^{\alpha}g_{0u}(\partial^{\mu}S^{\beta}_{B}) S_{A}^{\gamma} \partial^{\mu}S^{\delta}_{A}] . \end{equation}
The first term in this equation is proportional to the space derivative of the magnetic field $(\partial^{\mu}B^{\gamma})$.
It appears from the Zeeman energy (the first term in Hamiltonian (\ref{MFMafmUEI H Ham short HH and mB})),
like the first term in equation (\ref{MFMafmUEI P time derivative FM}) obtained for the ferromagnetic regime.
Other terms in equation (\ref{MFMafmUEI P time derivative AFM A}) contain
the interaction constants of the Heisenberg-Coulomb exchange interaction,
since their appearance is caused by this interaction from the second term in Hamiltonian (\ref{MFMafmUEI H Ham short HH and mB}).
The result for the second subspecies can be obtained via the exchange of subindexes $A \leftrightarrow B$.

It has been mentioned above that the polarization appears in the complex of two neighboring ions
that belong to different subspecies.
So, the partial polarization (\ref{MFMafmUEI P def AFM A}) and equation for its evolution (\ref{MFMafmUEI P time derivative AFM A}) are
the intermediate theoretical tools.
We need to combine the partial polarizations in the full polarization
$\partial_{t}P^{\alpha}=\partial_{t}P^{\alpha}_{A}+\partial_{t}P^{\alpha}_{B}$
and obtain the equation for its evolution
$$\partial_{t}P^{\alpha}
=-\frac{1}{3}\varepsilon^{\beta\gamma\delta}g_{\Pi}^{\alpha}\gamma (\partial^{\mu}B^{\gamma})
\partial^{\mu}(S_{A}^{\beta}S_{B}^{\delta})$$
$$+\frac{1}{3}\varepsilon^{\beta\gamma\delta}
[2g_{0\Pi}^{\alpha}g_{u}S^{\beta}_{B}S^{\gamma}_{A}\triangle (S_{A}^{\delta}-S_{B}^{\delta})$$
\begin{equation}\label{MFMafmUEI P time derivative AFM}
+g_{\Pi}^{\alpha}g_{0u}(\partial^{\mu}S^{\beta}_{B}) (S_{A}^{\gamma}-S_{B}^{\gamma}) \partial^{\mu}S^{\delta}_{A}] , \end{equation}
where the meaning of different terms is the same
as the physical meaning of the terms in equation (\ref{MFMafmUEI P time derivative AFM A}).

The spin evolution equations (\ref{MFMafmUEI s evolution HH AA and AB})
are combined in the evolution equations for functions
$\mbox{\boldmath $\Sigma$}=\textbf{S}_{A}+\textbf{S}_{B}$ and $\textbf{L}=\textbf{S}_{A}-\textbf{S}_{B}$,
which are traditionally used in the theory of the antiferromagnetic materials \cite{Landau vol 9}.
Therefore, it is essential to represent equation (\ref{MFMafmUEI P time derivative AFM}) in terms of these functions
$$\partial_{t}P^{\alpha}
=-\frac{1}{6}\varepsilon^{\beta\gamma\delta}g_{\Pi}^{\alpha}\gamma (\partial^{\mu}B^{\gamma})
\partial^{\mu}[L^{\beta}\Sigma^{\delta}]$$
\begin{equation}\label{MFMafmUEI P time derivative AFM LM}
+\frac{1}{6}\varepsilon^{\beta\gamma\delta}
[2g_{0\Pi}^{\alpha}g_{u}\Sigma^{\beta}L^{\gamma}\triangle L^{\delta}
+g_{\Pi}^{\alpha}g_{0u}(\partial^{\mu}\Sigma^{\beta}) L^{\gamma} \partial^{\mu}L^{\delta}] , \end{equation}
where we include
$\varepsilon^{\beta\gamma\delta}[(\Sigma^{\beta}+L^{\beta})(\Sigma^{\delta}-L^{\delta})]$
$=2\varepsilon^{\beta\gamma\delta}L^{\beta}\Sigma^{\delta}$.
Physical meaning of terms in the obtained equations can be traced via the coefficients.
The presence of the magnetic field shows the appearance of this term from the Zeeman energy,
and the presence of $g_{0u}$ or $g_{u}$ shows its appearance from the exchange interaction,
similarly to the equations shown above.

\section{Equilibrium solutions set of spin-polarization evolution equations}

Let us consider equilibrium structures obeying a system of equations for the spin evolution and polarization evolution.
For the simplicity of derivation of the equation obtained above,
we considered the external magnetic field.
It is possible to include the magnetic field created by the magnetic moments of the medium.
Hence, we need to include the Maxwell equations:
$\nabla\cdot\textbf{B}=0$
and
$\nabla\times\textbf{B}=4\pi\gamma\nabla\times\textbf{S}$,
where we included the zero time derivative of the electric field due to our focus on the static regime.

\subsection{Ferromagnetic multiferroics}

We start this analysis with the ferromagnetic materials.
In this case, we consider several configurations of the spin density.

\subsubsection{Parallel spins, transverse change of spin magnitude}

Let us consider the regime,
where we consider the spin density directed in the z-direction
$\textbf{S}_{0}=S_{0}\textbf{e}_{z}$.
We also assume that its module changes in the x-direction $S_{0}=S_{0}(x)$.
Since we assume $\partial_{t}\textbf{S}_{0}=0$,
we need to check
that the right-hand side of the spin evolution equation
\begin{equation}\label{MFMafmUEI Spin Evol F}
\partial_{t}\textbf{S}=\gamma[\textbf{S}\times\textbf{B}]
+\frac{1}{6}g_{u}[\textbf{S}\times\triangle\textbf{S}]\end{equation}
is equal to zero.
We see $\triangle\textbf{S}\parallel \textbf{e}_{z}$,
hence the last term is equal to zero.
To check the first term,
we need to find the corresponding magnetic field, assuming that the external magnetic field is equal to zero.
We find $\nabla\times\textbf{S}_{0}=-(\partial_{x}S_{0}(x))\textbf{e}_{y}$.
Hence, $\nabla\times\textbf{B}_{0}=-4\pi\gamma(\partial_{x}S_{0}(x))\textbf{e}_{y}$.
It gives $\textbf{B}_{0}=-4\pi\gamma S_{0}(x)\textbf{e}_{z}$.
The condition $\nabla\times \textbf{B}_{0}=0$ is also satisfied.
It is parallel to $\textbf{S}_{0}$,
so the first term in equation (\ref{MFMafmUEI Spin Evol F}) is equal to zero as well.

The equilibrium condition means that the polarization does not depend on time.
Hence, the right-hand side of equation (\ref{MFMafmUEI P time derivative FM}).
Estimations given above show that this condition is satisfied.
We also need to find the corresponding polarization (nonzero value)
via the first term on the right-hand side of equation (\ref{MFMafmUEI P def expanded FR}):
$\textbf{P}_{0}=\textbf{g}_{0\Pi}\textbf{S}_{0}^{2}$.
Let us remind
that constant $\textbf{g}_{0\Pi}$ is parallel to the shift of the ligand ion
$\mbox{\boldmath $\delta$}$.
We present a simple equilibrium spin structure leading to nonzero polarization.

\subsubsection{Parallel spins, longitudinal change of spin magnitude}

We consider the spin density directed in the z-direction
$\textbf{S}_{0}=S_{0}\textbf{e}_{z}$,
where its module changes in the z-direction $S_{0}=S_{0}(z)$ as well.
We require $\partial_{t}\textbf{S}_{0}=0$ and check the value of the right-hand side of the spin evolution equation (\ref{MFMafmUEI Spin Evol F}).
We see $\triangle\textbf{S}\parallel \textbf{e}_{z}$,
hence the last term is equal to zero.
We also find the zero magnetic field $\textbf{B}=0$.
Hence, both equations (\ref{MFMafmUEI P time derivative FM}) and (\ref{MFMafmUEI Spin Evol F}) are satisfied.
We also find the corresponding polarization $\textbf{P}_{0}(z)=\textbf{g}_{0\Pi}\textbf{S}_{0}^{2}(z)$.

\subsubsection{Cycloidal spiral spin structure}

Let us consider the spiral spin structure
that was earlier presented in works \cite{Mostovoy PRL 06}, \cite{Dong AinP 15}:
\begin{equation}\label{MFMafmUEI Spin spiral b direction}
\textbf{S}_{0}(\textbf{r})=s_{b}\textbf{e}_{y}\cos(\textbf{r}\cdot\textbf{q})
+s_{c}\textbf{e}_{z}\sin(\textbf{r}\cdot\textbf{q})
+s_{a}\textbf{e}_{x}, \end{equation}
where
$\textbf{q}=q\textbf{e}_{y}$.
It is a spiral shifting in the direction being in the rotation plane.
It can be represented in the following form
\begin{equation}\label{MFMafmUEI Spin spiral b direction simple}
\textbf{S}_{0}(y)=s_{y}\textbf{e}_{b}\cos(yq)
+s_{c}\textbf{e}_{z}\sin(yq)
+s_{a}\textbf{e}_{x}, \end{equation}
which can be substituted in the spin evolution equation to find the magnetic field corresponding to the equilibrium condition.

The right-hand side of equation (\ref{MFMafmUEI Spin Evol F}) should be equal to zero for the static regime.
In addition to the magnetic field parallel $\textbf{B}_{1}=\chi\textbf{S}_{0}$
to the equilibrium spin density,
we need to include an additional field
since the second term $[\textbf{S}_{0}\times\triangle\textbf{S}_{0}]$
$=-q^{2}[\textbf{S}_{0}\times(\textbf{S}_{0}-s_{a}\textbf{e}_{x})]$
$=s_{a}q^{2}[\textbf{S}_{0}\times\textbf{e}_{x}]$ has a nonzero value.
It leads to the following structure of the magnetic field
$\textbf{B}_{0}=\textbf{B}_{1}+\textbf{B}_{2}$
with the additional constant field
$\textbf{B}_{2}=-\frac{g_{u}\hbar}{12\gamma}q^{2}s_{a}\textbf{e}_{x}$.
To complete the solution,
we need to find the coefficient $\chi$.
If we assume $\chi=const$ we find that equation $\nabla\cdot\textbf{B}_{0}=0$ cannot be satisfied.
So we consider coefficient $\chi$ s a function of coordinates $\chi(\textbf{r})$.
However, the $x$ and $z$ projections of equation $\nabla\times\textbf{B}_{0}=4\pi\gamma\nabla\times\textbf{S}_{0}$
can be satisfied at $\chi=4\pi\gamma$ or $q=0$.
We conclude
that two interactions entering the spin evolution equation (\ref{MFMafmUEI Spin Evol F})
cannot support structure (\ref{MFMafmUEI Spin spiral b direction}).
Possibly, one can find a consistent solution in form (\ref{MFMafmUEI Spin spiral a direction simple})
by extending the range of interactions included in the model.

\subsubsection{Screw spiral spin structure}

Here we consider a spiral shifting in the direction perpendicular to the rotation plane,
so substitute $\textbf{q}=q\textbf{e}_{x}$ in equation (\ref{MFMafmUEI Spin spiral b direction}).
In the chosen regime, the structure simplifies to
\begin{equation}\label{MFMafmUEI Spin spiral a direction simple}
\textbf{S}_{0}(x)=s_{b}\textbf{e}_{y}\cos(xq)
+s_{c}\textbf{e}_{z}\sin(xq)
+s_{a}\textbf{e}_{x}. \end{equation}

Let us consider the right-hand side of equation (\ref{MFMafmUEI Spin Evol F})
under assumption (\ref{MFMafmUEI Spin spiral a direction simple}) for the spin structure.
The balance of two terms leads to the following form of magnetic field
$\textbf{B}_{0}=\textbf{B}_{1}+\textbf{B}_{2}$
with
$\textbf{B}_{1}=\chi\textbf{S}_{0}$
and
\begin{equation}\label{MFMafmUEI B 2}
\textbf{B}_{2}=-\frac{g_{u}\hbar}{12\gamma}q^{2}s_{a}\textbf{e}_{x}. \end{equation}

We need to check that the found magnetic field satisfies equation $\nabla\cdot\textbf{B}_{0}=0$.
It shows that function $\chi$ depends on coordinates $y$ and $z$ ($\chi(y,z)$) or to be a constant $\chi=const$.
Next, we need to consider the second Maxwell equation $\nabla\times\textbf{B}_{0}=4\pi\gamma\nabla\times\textbf{S}_{0}$ (its static regime).
Assuming $\chi=const$, we find the explicit expression for $\chi=4\pi\gamma$.

We can check
that the right-hand side of equation (\ref{MFMafmUEI P time derivative FM}) is equal to zero,
since we consider the equilibrium state.
We see that it is satisfied.
Let us present the corresponding polarization
$P_{0}^{\alpha}=g^{\alpha}_{0\Pi}\textbf{S}^{2}=g^{\alpha}_{0\Pi}(s_{a}^{2}+s_{b}^{2}\cos^{2}qx+s_{c}^{2}\sin^{2}qx)$.
It can be a constant under conditions $s_{b}=\pm s_{c}$.
It gives polarization $P_{0}^{\alpha}$ in the following form $P_{0}^{\alpha}=g^{\alpha}_{0\Pi}(s_{a}^{2}+s_{b}^{2})$.
Let us remind that
the constant $g^{\alpha}_{0\Pi}$ is parallel to the shift of the ligand ion $\delta^{\alpha}$
(\ref{MFMafmUEI Pi via beta}).

\subsection{Antiferromagnetic multiferroics}

We presented a spiral spin structure in the a-direction for ferromagnetic multiferroics
in terms of the model based on the Zeeman energy and the Heisenberg-Coulomb exchange interaction.
So, we are focused on the same regime for the antiferromagnetic multiferroics,
but we also briefly mention the uniform regime.

\subsubsection{Uniform regime}

For the uniform regime, we have parallel partial spin densities,
and therefore we have parallel vectors
$\textbf{L}_{0}$ and $\mbox{\boldmath $\Sigma$}_{0}$.
However, we can consider different modules of the partial spin densities in their opposite directions,
so $\mbox{\boldmath $\Sigma$}_{0}$ has a nonzero equilibrium value.
It corresponds to the constant magnetic field parallel to vectors $\textbf{L}_{0}$ and $\mbox{\boldmath $\Sigma$}_{0}$.

\subsubsection{On a form of screw spiral spin structure}

In the uniform case, we consider the parallel partial spin densities with different modules.
Here, we can consider two regimes of spirals in a-direction.
One corresponds to the parallel partial spin densities with different modules.
So, we see spirals for $\textbf{L}_{0}$ and $\mbox{\boldmath $\Sigma$}_{0}$ with the space phase shift on $\pi$.
Another case is the regime,
where the partial spin densities have approximately equal modules,
but they are directed at the angle to each other.
It leads to perpendicular directions of $\textbf{L}_{0}$ and $\mbox{\boldmath $\Sigma$}_{0}$ at each point.
It corresponds to the space phase shift on $\pi/2$ for $\textbf{L}_{0}$ and $\mbox{\boldmath $\Sigma$}_{0}$.

Let us start the analysis with the screw spiral structure for $\textbf{L}_{0}$ vector
\begin{equation}\label{MFMafmUEI Spin spiral AFM L a direction simple}
\textbf{L}_{0}(x)=l_{b}\textbf{e}_{y}\cos(xq)
+l_{c}\textbf{e}_{z}\sin(xq)
+l_{a}\textbf{e}_{x}, \end{equation}
while other characteristics
we retrieve from equilibrium regime of equations of motion.

Next, we need to find the magnetic field corresponding to both
the spin evolution equations (\ref{MFMafmUEI s evolution HH Sigma}), (\ref{MFMafmUEI s evolution HH L}),
and the Maxwell equations $\nabla\cdot\textbf{B}_{0}=0$ and $\nabla\times\textbf{B}_{0}=4\pi_{i=1}^{2}\nabla\times(\gamma_{i}\textbf{S}_{0i})$.
In the chosen case, we have $\gamma_{1}=\gamma_{2}$ and $\textbf{S}_{01}+\textbf{S}_{02}=\mbox{\boldmath $\Sigma$}_{0}$.

We consider the equilibrium form of equation (\ref{MFMafmUEI s evolution HH L}),
where we dropped the second term on the right-hand side.
It gives $(2\mu/\hbar)\textbf{B}_{0}=\chi \textbf{L}_{0}-g_{0u,AB}\mbox{\boldmath $\Sigma$}_{0}$,
where $\chi$ is an unknown coefficient.
We substitute this magnetic field in equation (\ref{MFMafmUEI s evolution HH Sigma})
and obtain vector $\mbox{\boldmath $\Sigma$}_{0}$:
\begin{equation}\label{MFMafmUEI Spin spiral AFM M a direction simple}
\mbox{\boldmath $\Sigma$}_{0}(x)=\frac{\alpha}{\chi}\textbf{L}_{0}+\frac{1}{6}g_{u}q^{2}l_{a}\textbf{e}_{x}, \end{equation}
where $\alpha$ is another unknown coefficient.
It also leads to the expression for the magnetic field
$\textbf{B}_{0}=(\hbar/2\mu)[(\chi- g_{0u,AB}\alpha/\chi)\textbf{L}_{0}-\frac{1}{6}g_{0u,AB}g_{u}q^{2}l_{a}\textbf{e}_{x}$.

Equation $\nabla\cdot\textbf{B}=0$ can be satisfied
if $(\chi- g_{0u,AB}\alpha/\chi)$ is a constant or a combination of functions equal to zero.
Equation
$\nabla\times\textbf{B}=4\pi\gamma\nabla\times\mbox{\boldmath $\Sigma$}_{0}$ can be satisfied
at the following relation between two introduced coefficients
\begin{equation}\label{MFMafmUEI chi via alpha} \chi^{2}=\alpha(4\pi\gamma\mu/\hbar+g_{0u,AB}).\end{equation}
Hence, $(\chi- g_{0u,AB}\alpha/\chi)$ is a nonzero constant.
So, $\alpha$ and $\chi$ are constants connected via equation (\ref{MFMafmUEI chi via alpha}).

To complete our analysis, we need to check
that the polarization evolution equation (\ref{MFMafmUEI P time derivative AFM LM}) also corresponds to the equilibrium regime,
so its right-hand side is equal to zero.
The direct substitution of found $\textbf{B}_{0}$, $\textbf{L}_{0}$, and $\mbox{\boldmath $\Sigma$}_{0}$
shows that it is satisfied.

In this case, the polarization is mostly defined by vector $\textbf{L}$
in accordance with equation (\ref{MFMafmUEI P def expanded AFR L M}).
If we need to get a constant value of polarization, we need to choose $l_{b}=\pm l_{c}$
and find
$\textbf{P}_{0}=(1/6)(\gamma/2mc)g_{(\beta)}\mbox{\boldmath $\delta$} (l_{a}^{2}+l_{b}^{2})$.

The spiral spin structures are the periodic magnetic structures,
which appears to be one of nontrivial spin structures along with
skyrmions, magnetic helix, magnetic vortex, chiral domain walls.
Spatial variation of the spin density is the key property for the polarization formation.
Hence, the spiral spin structure is one of structures which allows to obtain the electric polarization of the medium.
If we consider the antisymmetric form of the electric polarization following Ref. \cite{Mostovoy PRL 06},
the spiral spin structures are necessary for electric polarization in multiferroics.
The nonzero symmetric form of the electric polarization can be obtained
for the collinear spin density in accordance with equation (\ref{MFMafmUEI P DM from SCm 1}).
However, this is a formal result, since experimental estimation of the polarization is made up to normalization constant.
Nontrivial space dependence of the polarization can be found for spiral structures.
Moreover, the spiral spin structures allow to obtain the periodic change of the electric polarization in multiferroics.

\section{Conclusion}

The Landau--Lifshitz--Gilbert equation can be called
the main macroscopic equation for the evolution of the magnetization in the magnetically ordered materials.
Multiferroic materials show the existence of the electric polarization in addition to magnetization.
Hence, the study of the multiferroics requires a couple of connected equations for the magnetization and the electric polarization.
The problem of the derivation of described set of equations for the antiferromagnetic materials has been formulated in this paper.
The II-type of multiferroics with the electric dipole moment proportional to the scalar product of the neighboring spins
has been chosen for this research.
The polarization evolution equation has been found under the action of the Zeeman energy and the Heisenberg-Coulomb exchange interaction.
The similar equation for the ferromagnetic regime has been demonstrated as well.
The many-particle quantum hydrodynamic method has been applied for the derivation of the required polarization evolution equation.
Before, the application of this method to this derivation,
the method has been successfully tested on the derivation of the spin/magnetization evolution equation.

However, the chosen definition of the electric dipole moment has been required an analytical justification.
The justification has been made in several steps.
First, the spin-current model is justified for the ferromagnetic materials via the momentum balance equation
(the hydrodynamic Euler equation)
containing the spin-orbit interaction.
Second, the spin-current caused by the Dzylaoshinskii-Moriya interaction has been found from the spin/magnetization evolution equation
and placed in the spin-current model to find the required polarization.
Finally, the same steps have been made for the antiferromagnetic materials.

Therefore, it has been analytically derived
that there is
the electric dipole moment proportional to the scalar product of the neighboring spins
caused by the Dzylaoshinskii-Moriya.
The interpretation of the direction of the vector coefficient of proportionality in the electric dipole moment has been interpreted
as being parallel to the shift of the ligand ion from the line connecting neighboring magnetic ions
(this vector is the well-known part of the Dzylaoshinskii vector constant).

Some equilibrium spin configurations have been considered for the ferromagnetic and antiferromagnetic multiferroics.
Regimes of parallel and spiral spin structures have been discussed,
and corresponding electric polarizations have been calculated.

Overall, the Landau--Lifshitz--Gilbert equation contains the contribution of a number of physical mechanisms.
Their systematic account in the polarization evolution equation is the research program demonstrated in this paper.
The account of the Zeeman energy and the Heisenberg-Coulomb exchange interaction for the antiferromagnetic materials
with the electric dipole moment proportional to the scalar product of the neighboring spins
has been one of the initial steps towards the realization of this program.

\section{DATA AVAILABILITY}

Data sharing is not applicable to this article as no new data were
created or analyzed in this study, which is a purely theoretical one.

\section{Acknowledgements}
The work of M.I. Trukhanova is supported by the Russian Science Foundation under the grant No. 22-72-00036.

\end{document}